\documentclass[prc,amsmath,amssymb,showpacs,superscriptaddress,twocolumn]
{revtex4}
\usepackage{dcolumn,epsfig,color}

\begin{document} 

\title{Evolution of low-lying M1 modes in germanium isotopes}

\author{S. Frauendorf}\email{sfrauend@nd.edu}
\affiliation{Department of Physics, University of Notre Dame,
             Notre Dame, Indiana 46556, USA}

\author{R. Schwengner} 
\affiliation{Helmholtz-Zentrum Dresden-Rossendorf, 01328 Dresden, Germany}

\date{\today}

\begin{abstract}

Magnetic dipole strength functions are determined for the series of germanium
isotopes from $N = Z = 32$ to $N$ = 48 on the basis of a large number of
transition strengths calculated within the shell model. The evolution of the
strength with increasing neutron number in the $1g_{9/2}$ orbital is analyzed.
A bimodal structure comprising an enhancement toward low transition energy and
a resonance in the region of the scissors mode is identified. The low-energy
enhancement is strongest near closed shells, in particular at the almost
completely filled $1g_{9/2}$ orbital, while the scissorslike resonance is most
pronounced in the middle of the open shell, which correlates with the
magnitude of the also deduced electric quadrupole transition strengths. The
results are consistent with previous findings for the shorter series of iron
isotopes and prove the occurrence and correlation of the two low-lying
magnetic dipole modes as a global structural feature.
  
\end{abstract}


\maketitle

\section{Introduction}
\label{sec:intro}

The excitation and deexcitation of the nucleus by electromagnetic radiation at
high excitation energy and high level density are described by means of
$\gamma$-ray strength functions ($\gamma$-SF) which represent average
transition strengths in a certain energy range. The experimental determination
and the theoretical understanding of the properties of $\gamma$-SF has
attracted increasing interest because of their importance for the accurate
description of photonuclear reactions and the inverse radiative-capture
reactions, which play a central role in in the synthesis of the elements in
various stellar environments \cite{arn07,kap11}. The region of low transition
energies is important for these processes. Traditionally, the $\gamma$-SF has
been associated with electric dipole ($E1$) transitions.  Recently a strong
component of $M1$ character has been observed as an upbend of the  $\gamma$-SF
toward transition energy $E_\gamma = 0$.

First observed in $^{56,57}$Fe \cite{voi04}, the upbend was found also in
various nuclides in other mass regions, such as in Mo isotopes \cite{gut05},
in $^{105,106}$Cd \cite{lar13}, and in Sm isotopes \cite{sim16,sim19}. The
experiments used light-ion induced reactions such as ($^3$He,$^3$He'), and the
data were analyzed with the so-called Oslo method to extract level densities
and $\gamma$-SFs. This method was also applied in connection
with $\beta$ decay to $^{76}$Ga \cite{spy14}. A dominant dipole character of
the low-energy strength was demonstrated in Ref.~\cite{lar13L}, and an
indication for a magnetic dipole ($M1$) character was discussed for the case of
$^{60}$Ni \cite{voi10}.

Shell-model calculations revealed that a large number of $M1$ transitions
between excited states produces an exponential increase of the $\gamma$-ray
strength function that peaks at $E_\gamma \approx 0$ and describes the
low-energy enhancement of dipole strength observed in Mo isotopes around the
neutron shell closure at $N$ = 50 \cite{sch13L}. In these calculations, large
reduced transition strengths $B(M1)$ appear for transitions linking states with
configurations dominated by both protons and neutrons in high-$j$ orbitals,
the spins of which recouple. The low-energy enhancement was confirmed in
shell-model calculations for $^{56,57}$Fe \cite{bro14Fe},
$^{46,50,54}$Ti \cite{sie17a} and $^{44}$Sc \cite{sie17}. In the latter work,
also the electric dipole ($E1$) strength function was calculated, which does
not show an upbend comparable to that of the $M1$ strength. A correlation
between the low-energy $M1$ strength (LEMAR - Low Energy Magnetic Radiation) 
and the scissors resonance (SR), a fundamental $M1$ excitation occurring in
deformed nuclei around 3 MeV \cite{hey10}, was found in shell-model
calculations for the series of isotopes from $^{60}$Fe to $^{68}$Fe
\cite{sch17}. It was found that the low-energy $M1$ strength decreases and the
scissors strength develops when going into the open shell. The simultaneous
appearance of the two $M1$ modes is in accordance with experimental findings in
Sm isotopes \cite{sim16,sim19}. Later on, $M1$ strength functions were 
calculated for isotopic series in various mass regions \cite{kar17,sie18,mid18}.
The study in Ref.~\cite{mid18} confirmed that the low-energy $M1$ strength is
strongest in nuclides near shell closures.

In the present work we study the low-energy $\gamma$-SFs for the chain of the
Ge isotopes. The relatively small configuration space allows us to carry out
shell model calculations covering completely the open neutron shell
$32\leq N\leq 48$. We demonstrate for the first time that the low-energy $M1$
strength is concentrated in the LEMAR spike at the bottom of the shell, it is
partially moved into the SR in the middle of the shell, and is again
concentrated in the LEMAR spike at the top of the shell.

\section{Shell-model calculations}
\label{sec:shell}

The shell-model calculations for the germanium isotopes were carried out in the
jj44pn model space with the jj44bpn Hamiltonian \cite{hon09,bro,lis04} using
the code NuShellX@MSU \cite{bro14Nu}. The model space included the proton
and neutron orbitals $(1f_{5/2}, 2p_{3/2}, 2p_{1/2}, 1g_{9/2})$.
At first, we calculated the energies of the yrast states and the reduced
transition strengths of the linking electric quadrupole ($E2$) transitions for
varied limitations of occupation numbers to test at which numbers the $B(E2)$
values do not change further and a convergence is achieved. This is in
particular important for the mid-shell isotopes. For example, an increase of
the allowed maximum occupation number (upper limit) in the neutron $1g_{9/2}$
orbital from four to six in $^{70}$Ge does not change the
$B(E2, 2^+_1 \rightarrow 0^+_1)$ value and, thus, the application of an upper
limit of four is appropriate. In $^{74}$Ge, a change of this number from six to
eight changes the $B(E2, 2^+_1 \rightarrow 0^+_1)$ value from 357 to
369 e$^2$fm$^4$, while the further increase to ten neutrons does not cause any
further change. At the same time, the allowed minimum occupation numbers
(lower limits) in the neutron $2p_{3/2}$ and $1f_{5/2}$ orbitals were set to two.
A decrease of these lower limits to zero results in 
$B(E2, 2^+_1 \rightarrow 0^+_1)$ = 371 e$^2$fm$^4$.
In the full calculations including all transition strengths, the following
limits of occupation numbers were applied to truncate the configuration space
and, hence, make the calculations feasible and efficient. Up to four protons
were allowed to occupy each of the $1f_{5/2}$ and $2p_{3/2}$ orbitals while up
to two could be lifted to each of the $2p_{1/2}$ and $1g_{9/2}$ orbitals. The
same holds for the neutrons in $^{64}$Ge, while there can be up to six neutrons
in the $1f_{5/2}$ orbital in $^{66}$Ge. In $^{70}$Ge, at least two neutrons are
in each of the $1f_{5/2}$ and $2p_{3/2}$ orbitals and up to four can be excited
to the $1g_{9/2}$ orbital. The possible occupation numbers of neutrons in the
$1g_{9/2}$ orbital are further increased in the heavier isotopes, ranging from
two to eight in $^{74}$Ge, from six to ten in $^{78}$Ge, and from eight to ten
in $^{80}$Ge. 
For the calculation of the reduced electric quadrupole transition
strengths $B(E2)$, standard effective charges of $e_\pi = 1.5 e$ and
$e_\nu = 0.5 e$ were used and for the $B(M1)$ strengths, effective $g$ factors
of $g^{\rm eff}_s = 0.7  g^{\rm free}_s$ were applied.

The full calculations were performed for the lowest 40 states of each spin from
$J_i, J_f$ = 0 to 10 and each parity. The reduced transition strengths $B(M1)$
were calculated for all transitions from initial to final states with energies
$E_f < E_i$ and spins $J_f = J_i, J_i \pm 1$. This resulted in more than 24000
$M1$ transitions for each parity, which were sorted into 0.1 MeV bins of 
transition energy $E_\gamma = E_i - E_f$. The average $B(M1)$ value for one
energy bin was obtained as the sum of all $B(M1)$ values divided by the number
of transitions within this bin. Average $B(E2)$ values were deduced in an
analogous way, but include also the $\Delta J$ = 2 transitions. 
$M1$ strength functions were deduced according to

\begin{multline}
\label{eq:f1M1}
f_{M1}(E_\gamma,E_i,J_i,\pi) =  \\
16\pi/9~(\hbar c)^{-3}~\overline{B}(M1,E_i \rightarrow E_f, J_i, \pi)~\rho(E_i, J_i, \pi),
\end{multline}

where the $\overline{B}(M1,E_i \rightarrow E_f,J_i,\pi)$ are averages in
considered $(E_i,E_f)$ bins for given $J_i$, $\pi$, and $\rho(E_i,J_i,\pi)$ are
the level densities derived from the present calculations. The strength
functions $f_{M1}(E_\gamma)$ were obtained by averaging step-by-step over
$E_i$, $J_i$, and $\pi$.

\section{Results for the yrast region}
\label{sec:yrast}

\begin{figure}
\epsfig{file=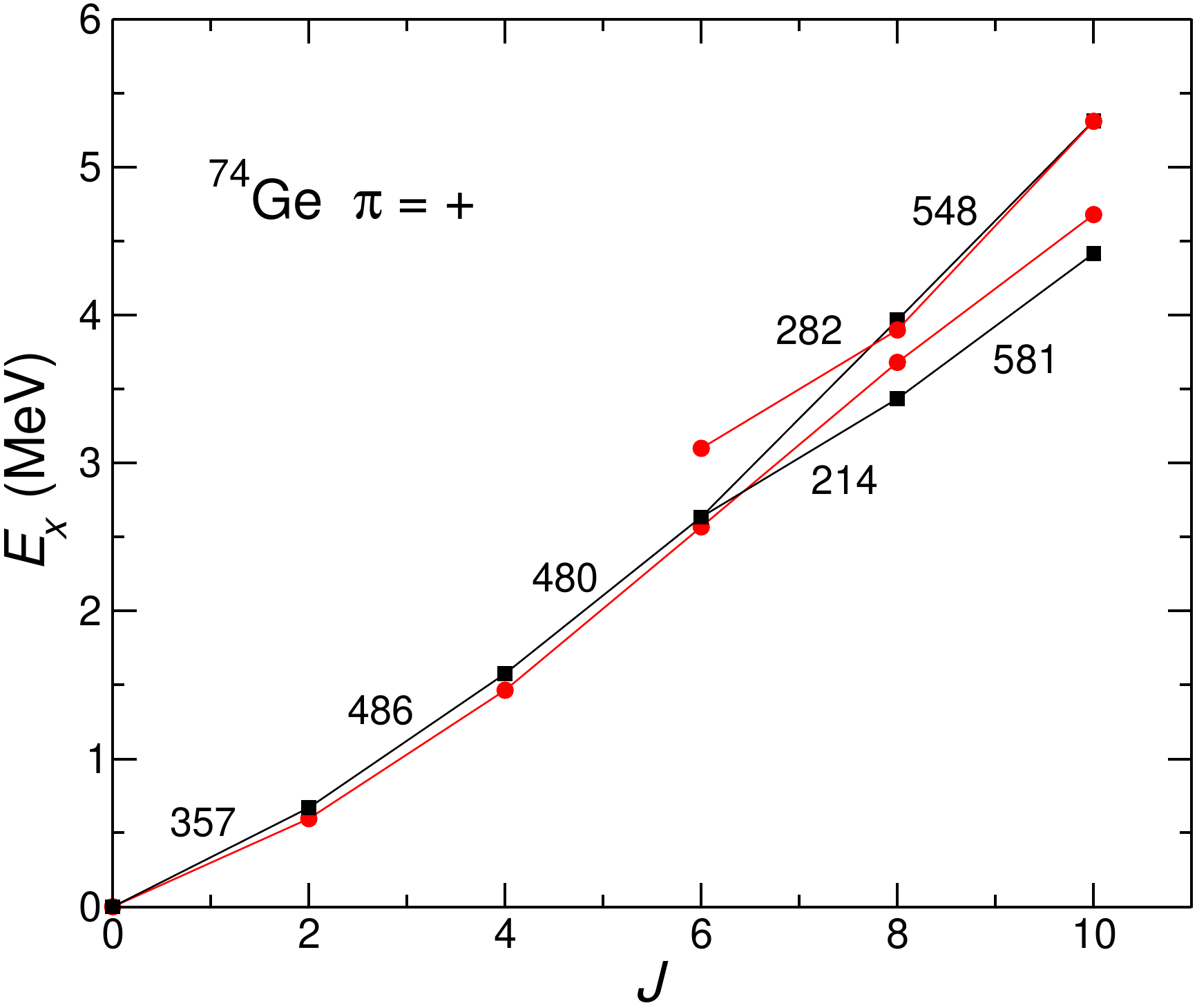,width=5.8cm}  
\caption{\label{fig:74Ge6JE} Experimental (red circles) and calculated (black
  squares) excitation energies of the yrast states and the states of the first
  excited band in $^{74}$Ge. The lines represent the linking $E2$ transitions
  with large strengths. The numbers at the lines are calculated $B(E2)$ values
  in e$^2$fm$^4$. The configuration space allowed up to six neutrons in the
  $1g_{9/2}$ orbital (see text).}
\end{figure}

\begin{figure}
\epsfig{file=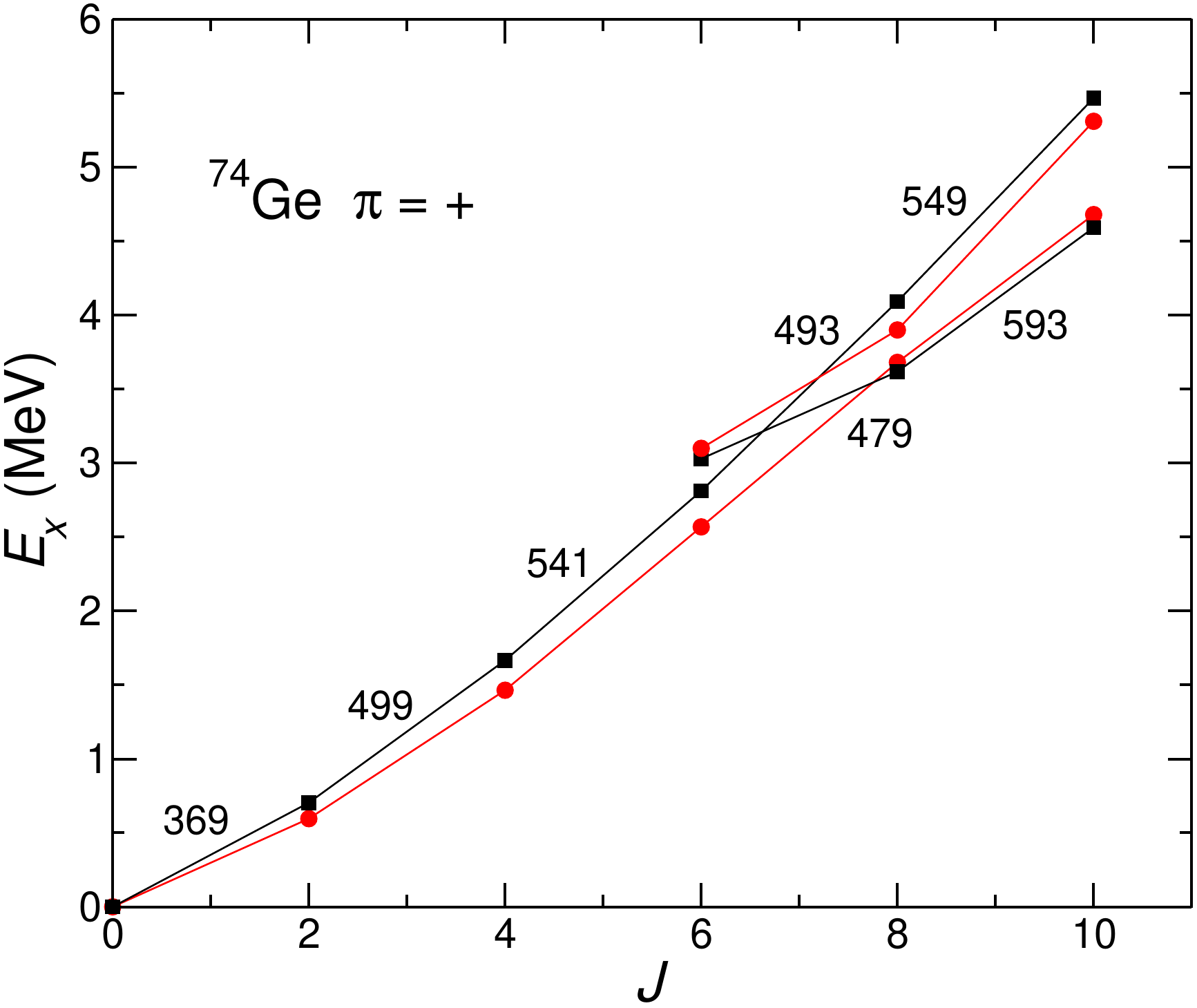,width=5.8cm}  
\caption{\label{fig:74Ge8JE} As Fig.~\ref{fig:74Ge6JE}, but with a
  configuration space allowing up to eight neutrons in the $1g_{9/2}$ orbital
  (see text).}
\end{figure}

To check the reliability of the shell-model calculations, we studied the yrast
regions of the Ge isotopes.  The calculated energies of the ground-state and
first excited bands in $^{74}$Ge are compared with the experimental ones
\cite{sun14} in Figs.~\ref{fig:74Ge6JE} and \ref{fig:74Ge8JE}. 
They represent the results obtained with the just discussed upper limits of six
and eight $1g_{9/2}$ neutrons, respectively. In both cases, the experimental
bands are well described by the calculations. A similarly good description of
the experimental yrast and yrare bands by the present calculations is achieved
for all other isotopes, which are presented in Figs.~\ref{fig:64,66,70GeJE} and
\ref{fig:78,80GeJE} of the appendix.

In accordance with the experiment, the energies $E(J)$ increase in a regular
way, forming quasi-rotatonal bands. The crossing of the ground-state band with
an excited band at $J = 6$ is reproduced. The energy ratios $E(J)/E(2^+)$
deviate substantially from the rotor rule $J(J + 1)$, The calculated ratios
$E(4^+_1)/E(2^+_1)$ of 2.44, 2.59, 2.59, 2.38, 2.32, 2.22 for
$^{64,66,70,74,78,80}$Ge 
compare well with experimental ratios of 2.28, 2.27,
2.07, 2.46, 2.54, 2.64, respectively. They characterize the Ge isotopes as soft
nuclei in the transitional region between spherical and deformed shapes,
because they are well below the rotor ratio of 3.33. The calculated $B(E2)$
values increase toward high spin up to the crossing region. For $^{74}$Ge, the
experimental ratio
$B(E2, 4^+_1 \rightarrow 2^+_1)/B(E2, 2^+_1 \rightarrow 0^+_1) = 1.33(8)$ is
reproduced by the calculated one of 1.35. 

The results for the $2^+_1$ states are compared with the experimental values in
Table~\ref{tab:BE2}.  The energies of the $2^+_1$ states are reproduced within
100 keV, except the  high experimental value for $^{70}$Ge. The experimental
$B(E2, 2^+_1 \rightarrow 0^+_1)$ values indicate a maximum of the quadrupole
collectivity in the middle of the shell, which is reproduced by the
calculations. However, the calculated peak is much shallower than in the
experiment. A similar shallow peak is obtained  for $J = 4$ and 6 (see
Figs.\ref{fig:64,66,70GeJE} and \ref{fig:78,80GeJE} of the appendix.).

One should be aware that the determination of the quadrupole collectivity from
the  $B(E2, 2^+_1 \rightarrow 0^+_1)$ values only is based on the assumption of
a rotational behavior of the yrast states, which is not realized for the soft
nuclei under consideration. Instead, the sums of the
$B(E2, 0^+_1 \rightarrow 2^+_n)$ values of all transitions from the ground
state are more appropriate \cite{Kum72,pov20} and also given in
Table~\ref{tab:BE2}. These sums of all $E2$ transitions into the ground state
are however only little larger and follow the trends of the
$B(E2, 2^+_1 \rightarrow 0^+_1)$ values. An even more comprehensive indicator
of the collectivity may be the consideration of average $B(E2)$ values between
all the states considered here. The further discussion of $B(E2)$ values in
Sec.~\ref{sec:res} therefore takes into account these values.

\begin{table}
\caption{\label{tab:BE2}Experimental and calculated reduced $E2$ transition
probabilities for the $2^+_1 \rightarrow 0^+_1$ transitions in
$^{64,66,70,74,78,80}$Ge.}
\begin{ruledtabular}
\begin{tabular}{cccccc} 
   & \multicolumn{2}{c}{$E(2^+_1$)} & 
  \multicolumn{2}{c}{$B(E2, 2^+_1 \rightarrow 0^+_1$)} &
  $\Sigma B(E2, 2^+_n \rightarrow 0^+_1)$                           \\
  & \multicolumn{2}{c}{(keV)} & \multicolumn{2}{c}{(e$^2$fm$^4$)} &
                                                   (e$^2$fm$^4$)    \\
     \cline{2-3} \cline{4-5} \cline{6-6}\\
  & EXP\footnotemark[1]  & CALC & EXP\footnotemark[1] & CALC & CALC \\
\hline
$^{64}$Ge$_{32}$ &  902 &  883 &         & 296 & 308 \\
$^{66}$Ge$_{34}$ &  957 &  828 & 190(36) & 300 & 321 \\
$^{70}$Ge$_{38}$ & 1040 &  649 & 356(7)  & 336 & 382 \\
$^{74}$Ge$_{42}$ &  596 &  704 & 609(7)  & 369 & 384 \\
$^{78}$Ge$_{46}$ &  619 &  782 & 455(79) & 320 & 343 \\
$^{80}$Ge$_{48}$ &  659 &  871 & 279(55) & 233 & 280 \\
\end{tabular}
\end{ruledtabular}
\footnotetext[1]{The values for $^{64,66,70,74,78,80}$Ge were taken from 
Refs.~\cite{nds64,nds66,nds70,nds74,nds78,nds80}, respectively.}
\end{table}

\begin{figure*}
\epsfig{file=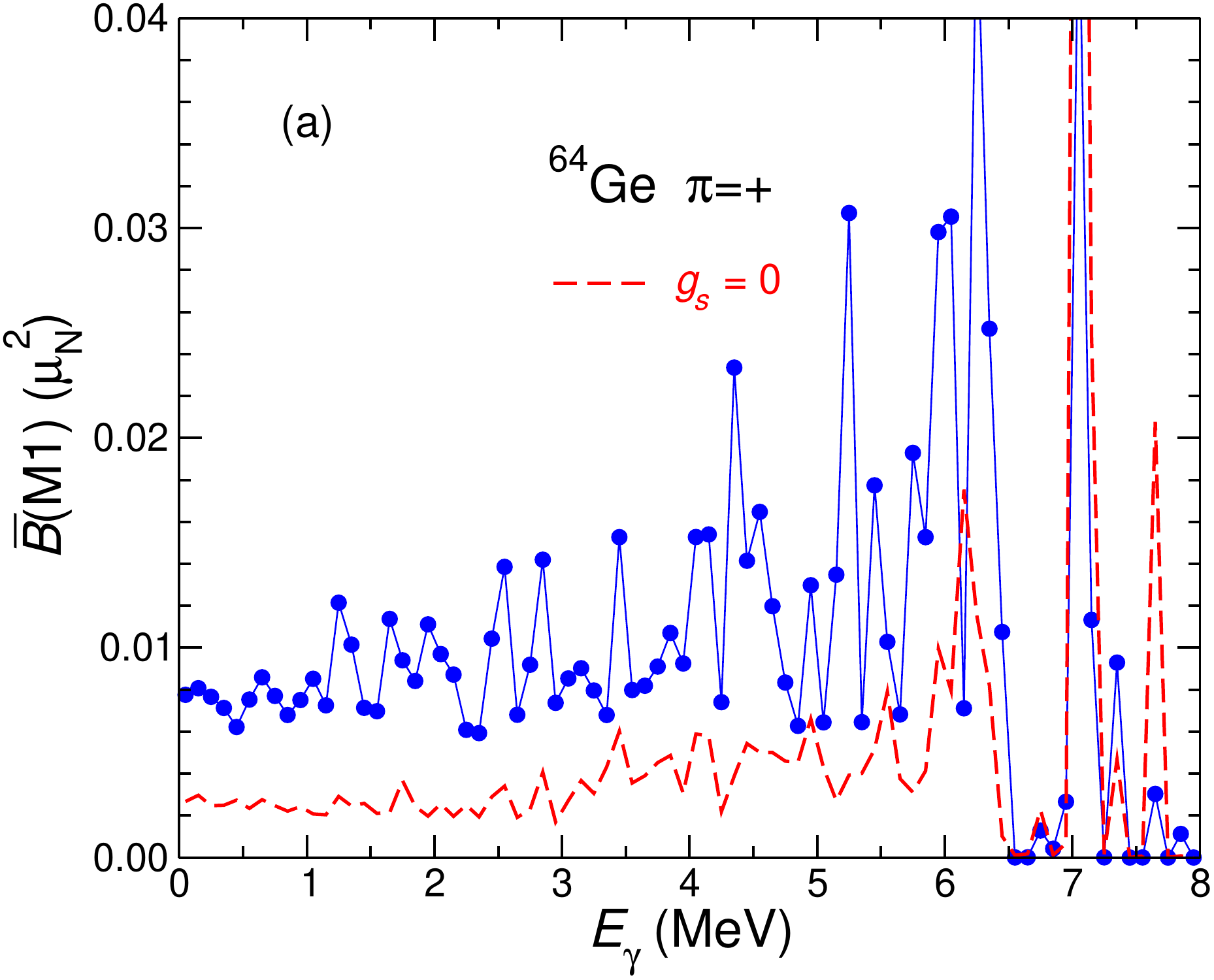,width=5.8cm}
\epsfig{file=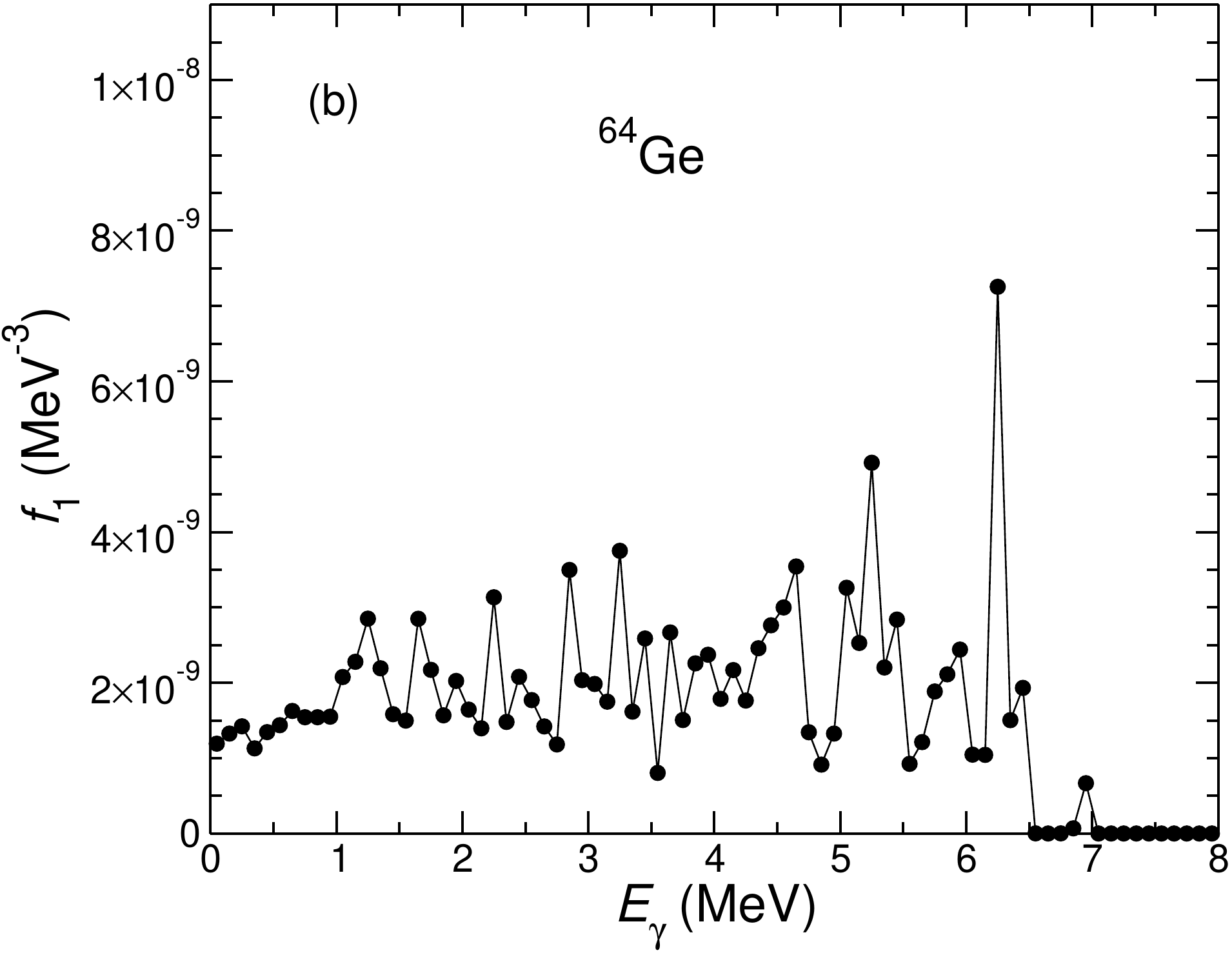,width=5.9cm}
\epsfig{file=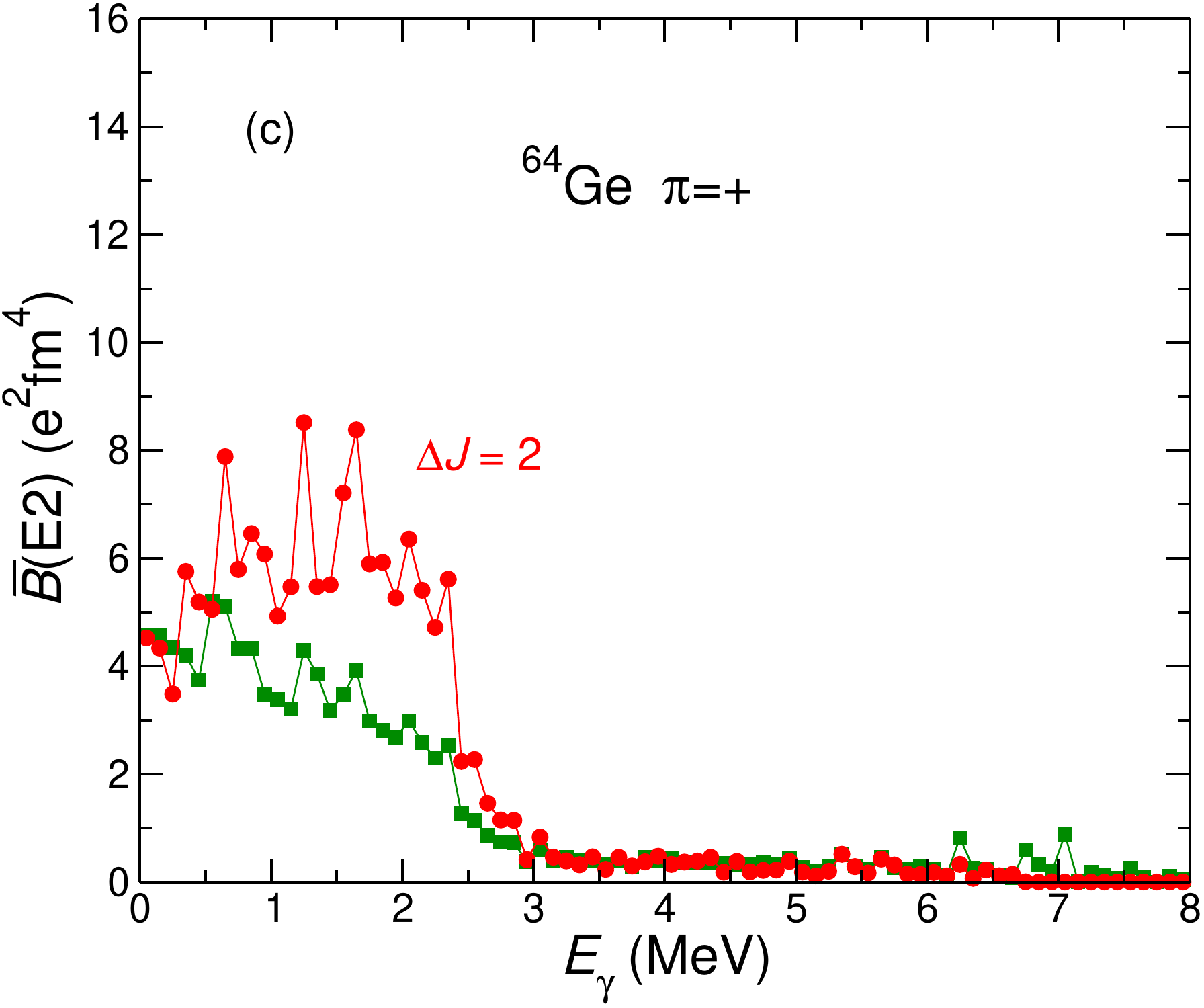,width=5.6cm}  
\caption{\label{fig:A64} Results for $^{64}$Ge in energy bins of 0.1 MeV.
  (a): Average reduced $M1$ transition strengths (blue circles) and
  their orbital contributions ($g_s$ = 0) only (red dashed line) for
  positive-parity states.
  (b): The $M1$ strength function including both parities.
  (c): Average reduced $E2$ transition strengths (green squares) and
  average $B(E2)$ values for stretched transitions with $J_i = J_f + 2$
  only (red circles) for positive-parity states.}
\end{figure*}

\begin{figure*}
\epsfig{file=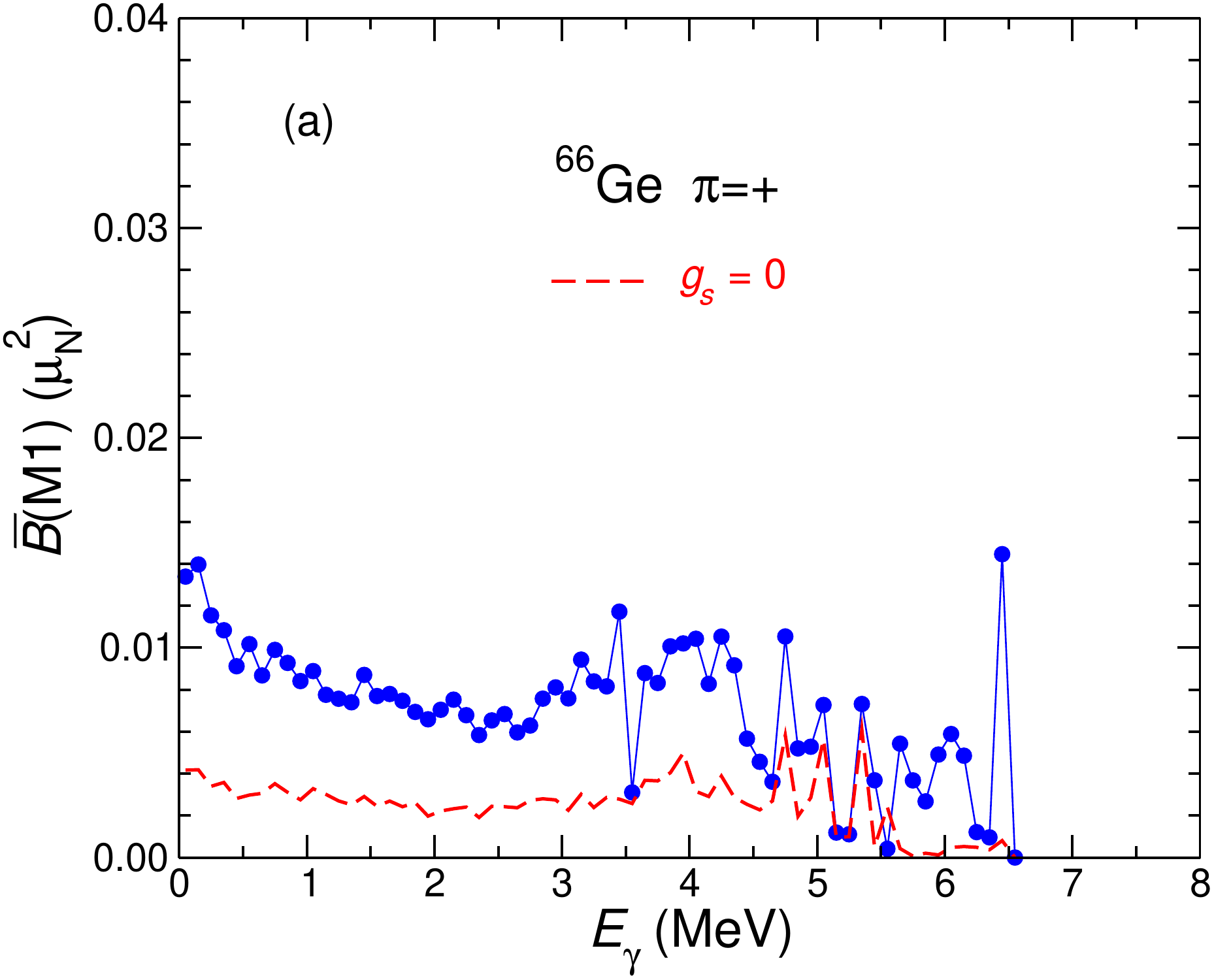,width=5.8cm}
\epsfig{file=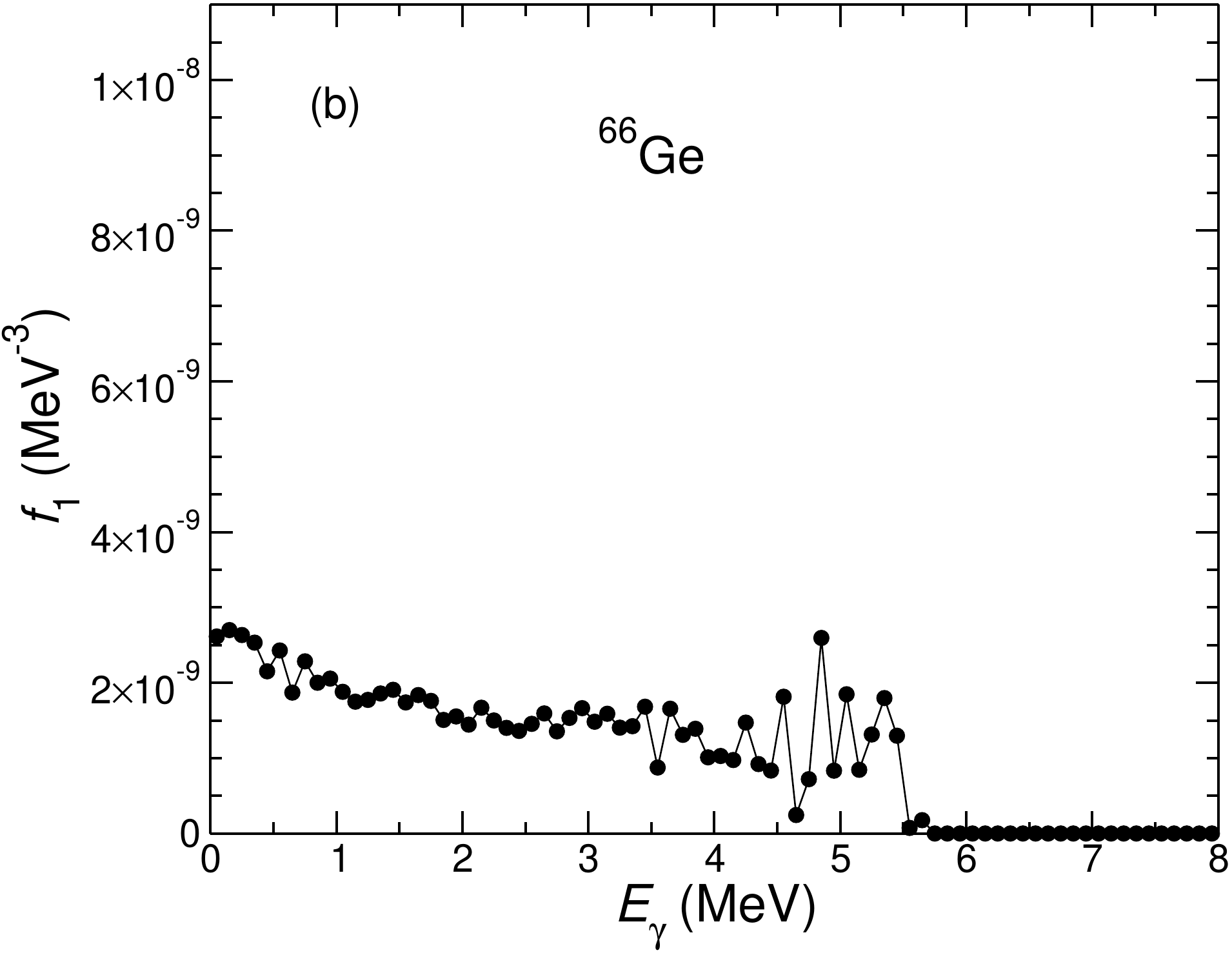,width=5.9cm}
\epsfig{file=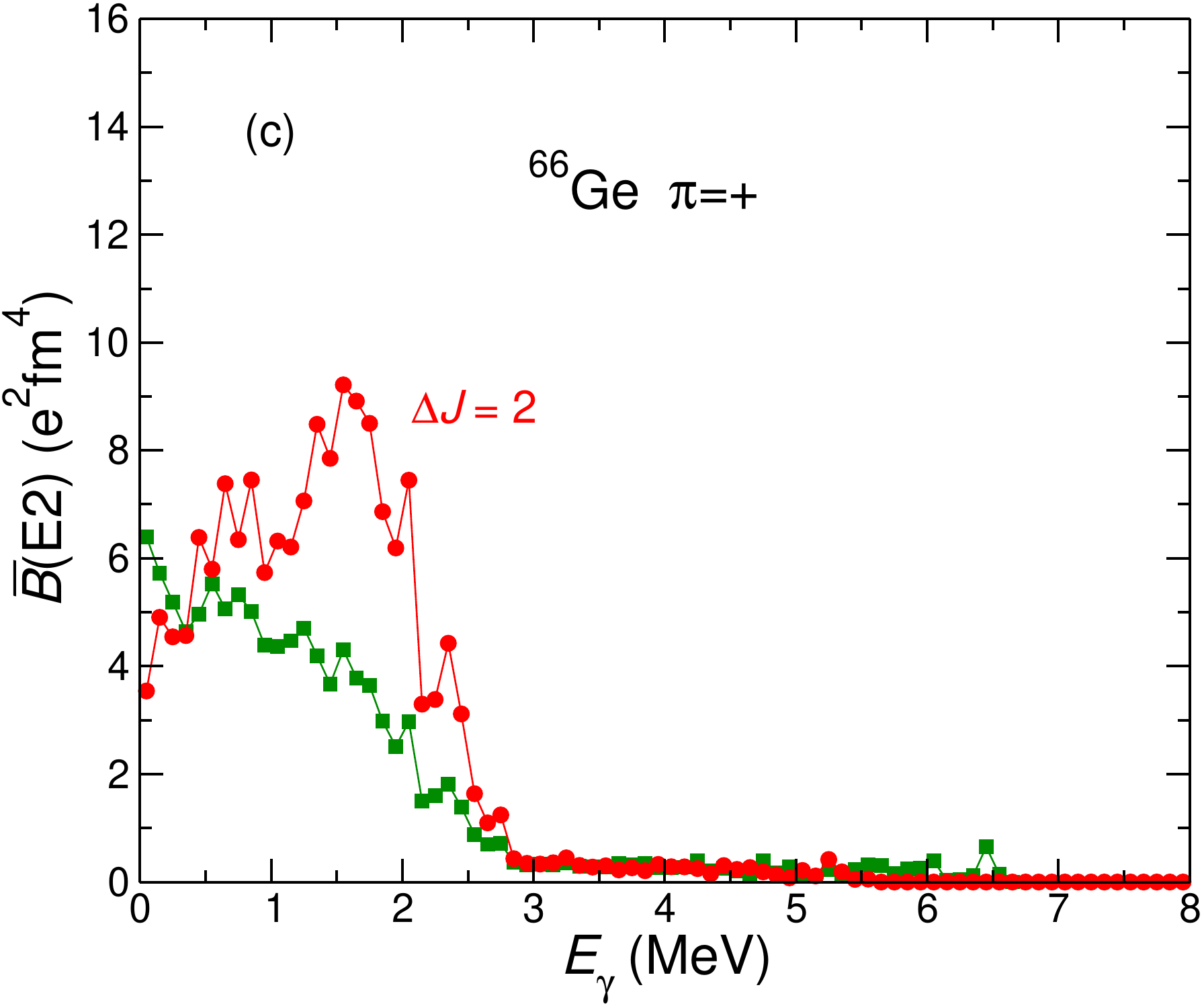,width=5.6cm}  
\caption{\label{fig:A66} As Fig.~\ref{fig:A64}, but for $^{66}$Ge.}
\end{figure*}

\begin{figure*}
\epsfig{file=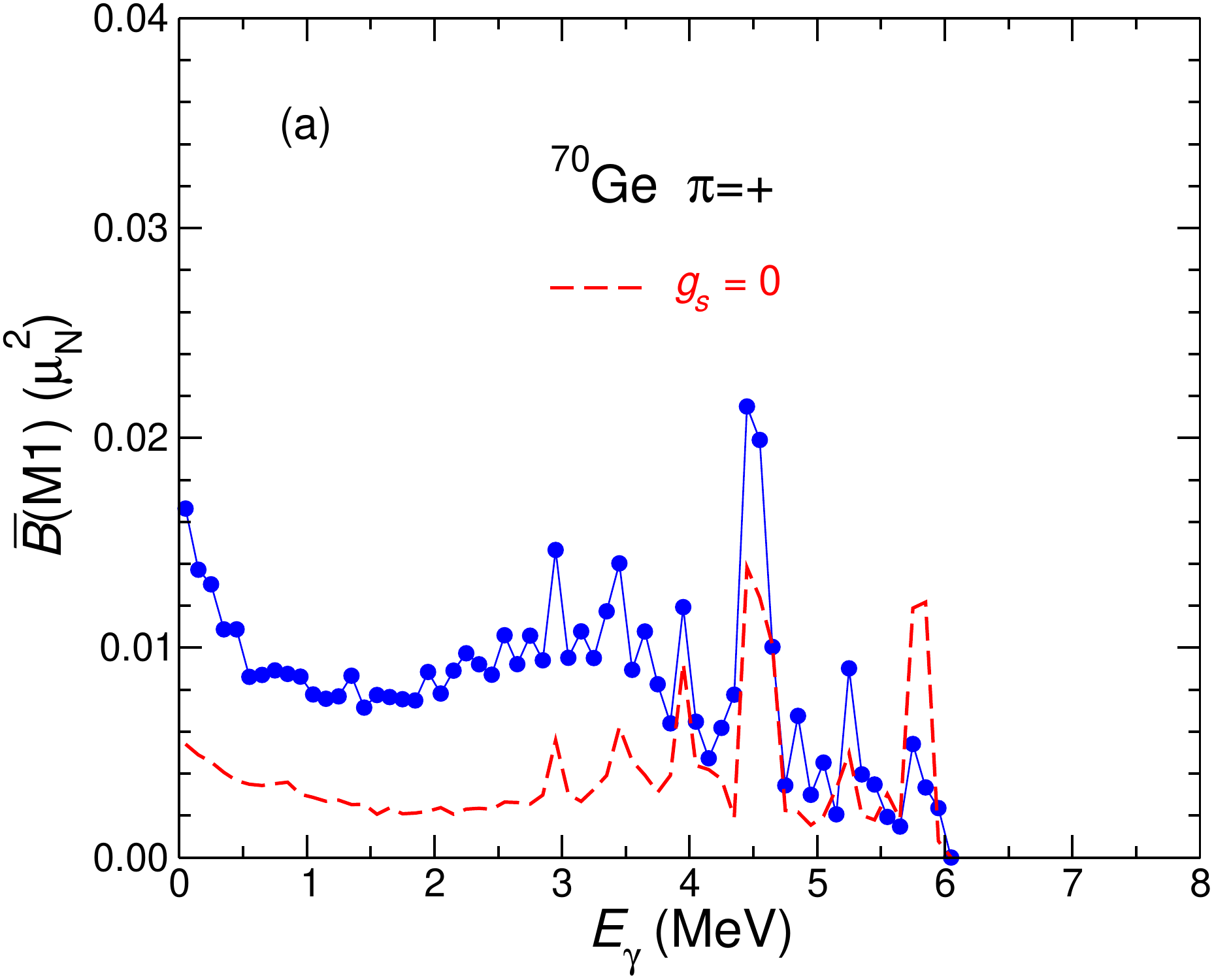,width=5.8cm}
\epsfig{file=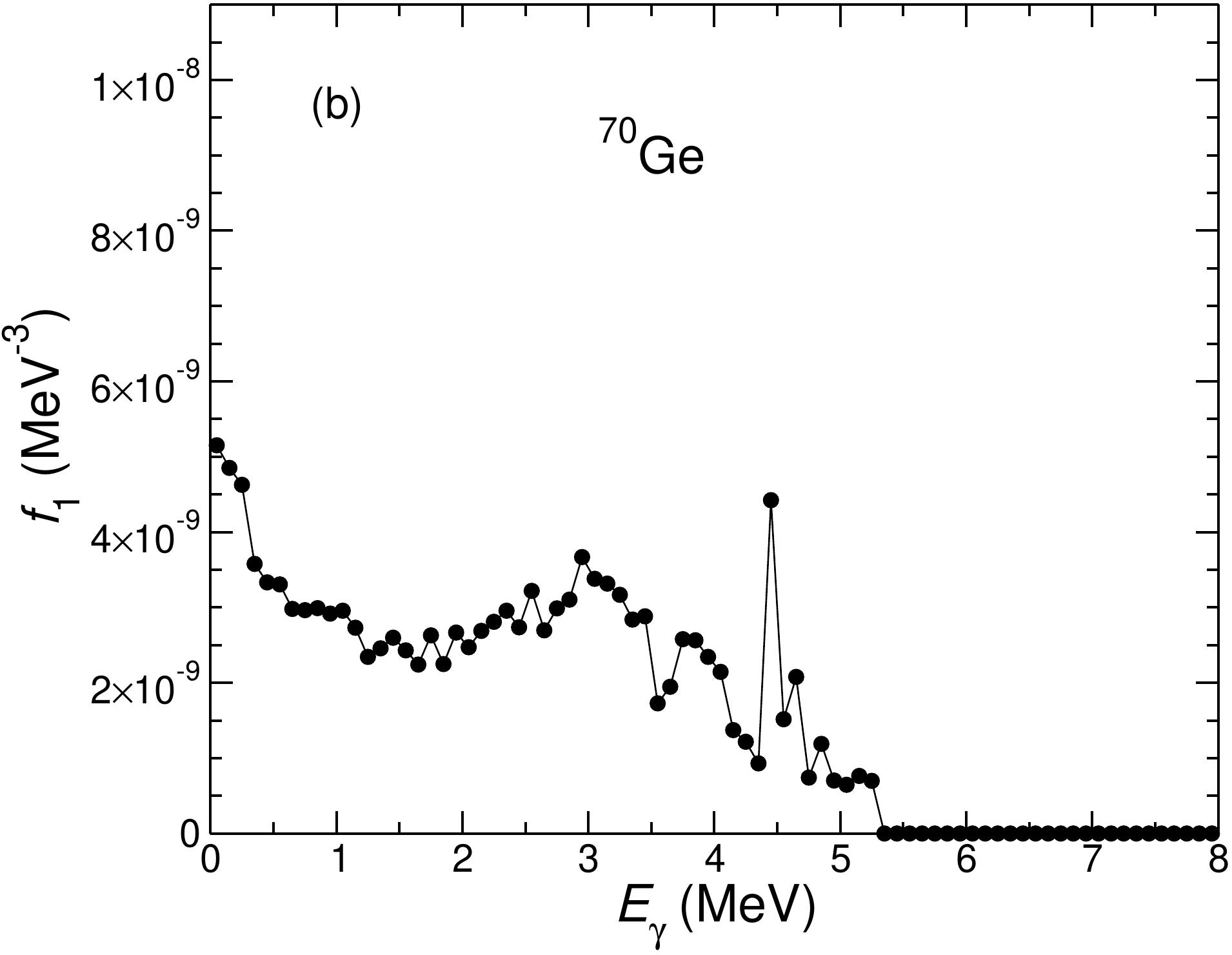,width=5.9cm}
\epsfig{file=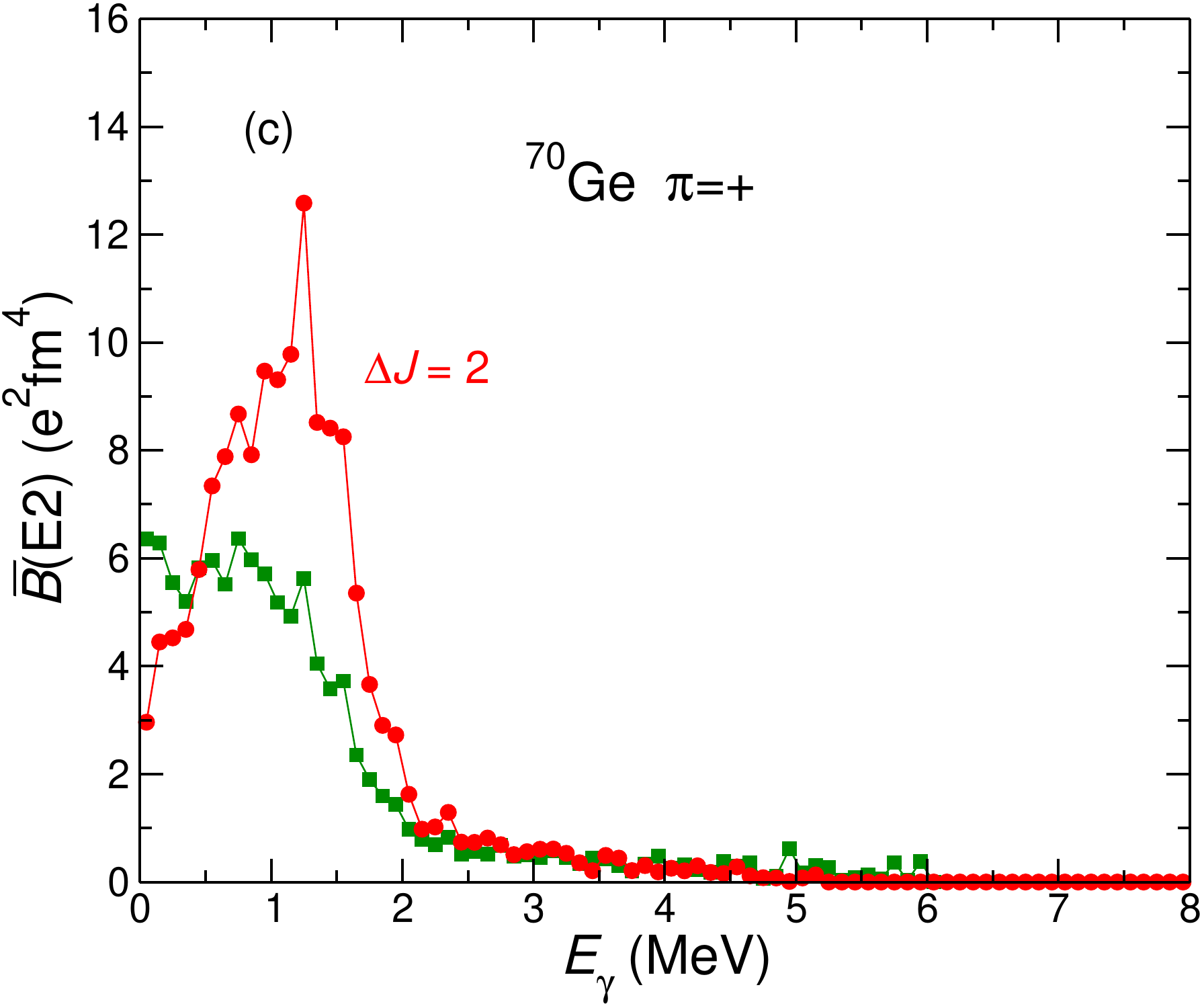,width=5.6cm}  
\caption{\label{fig:A70} As Fig.~\ref{fig:A64}, but for $^{70}$Ge.}
\end{figure*}
  
\begin{figure*}
\epsfig{file=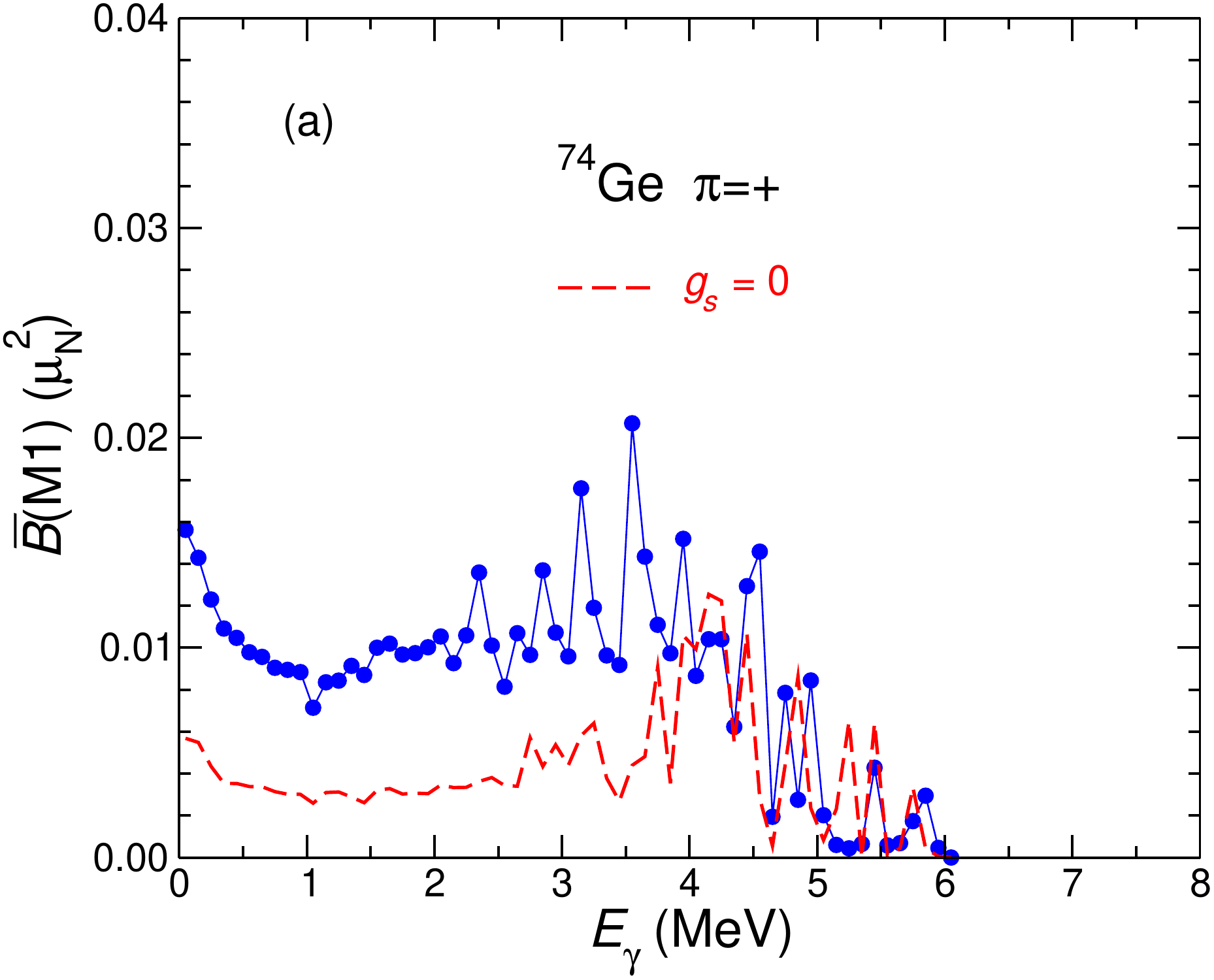,width=5.8cm}
\epsfig{file=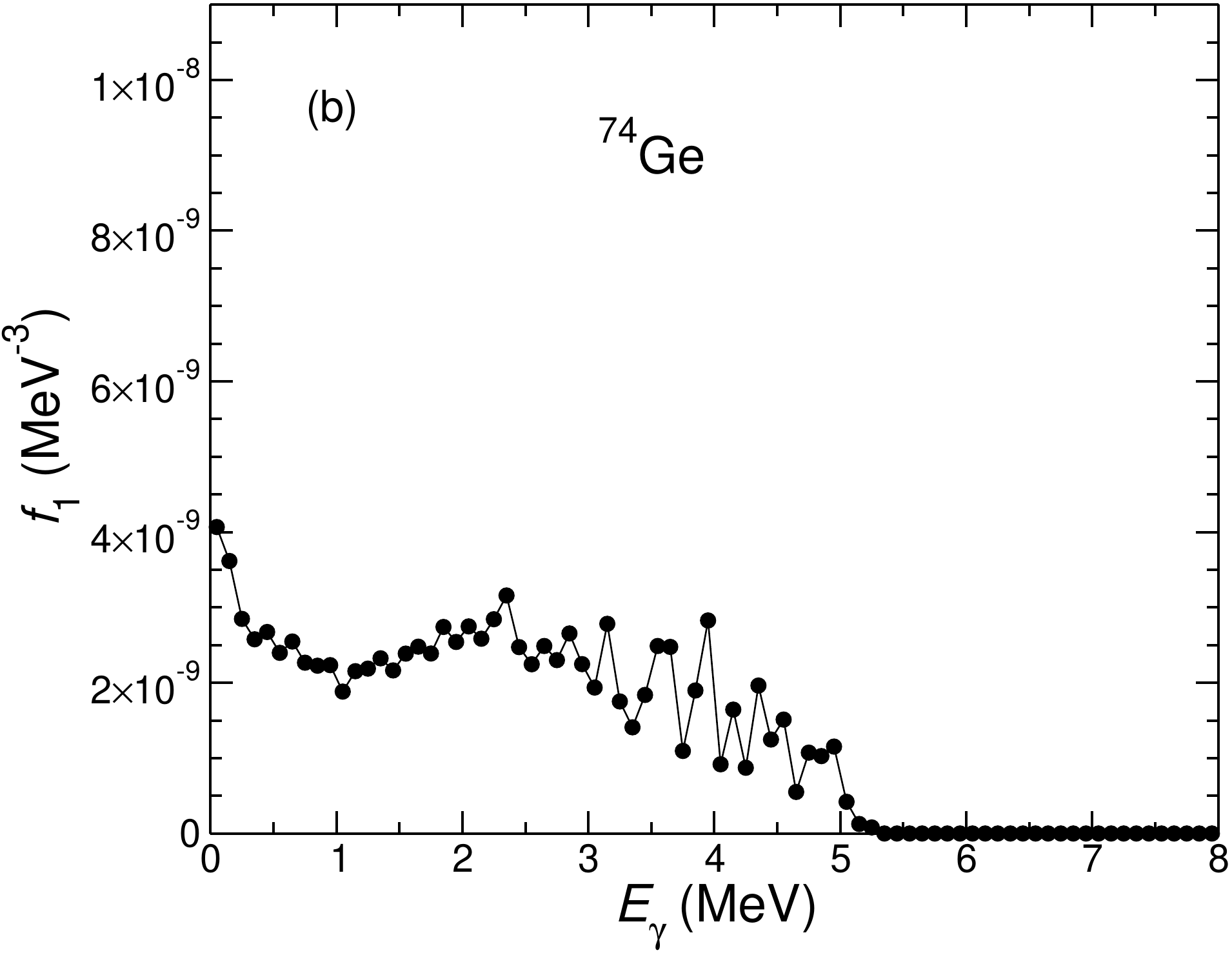,width=5.9cm}
\epsfig{file=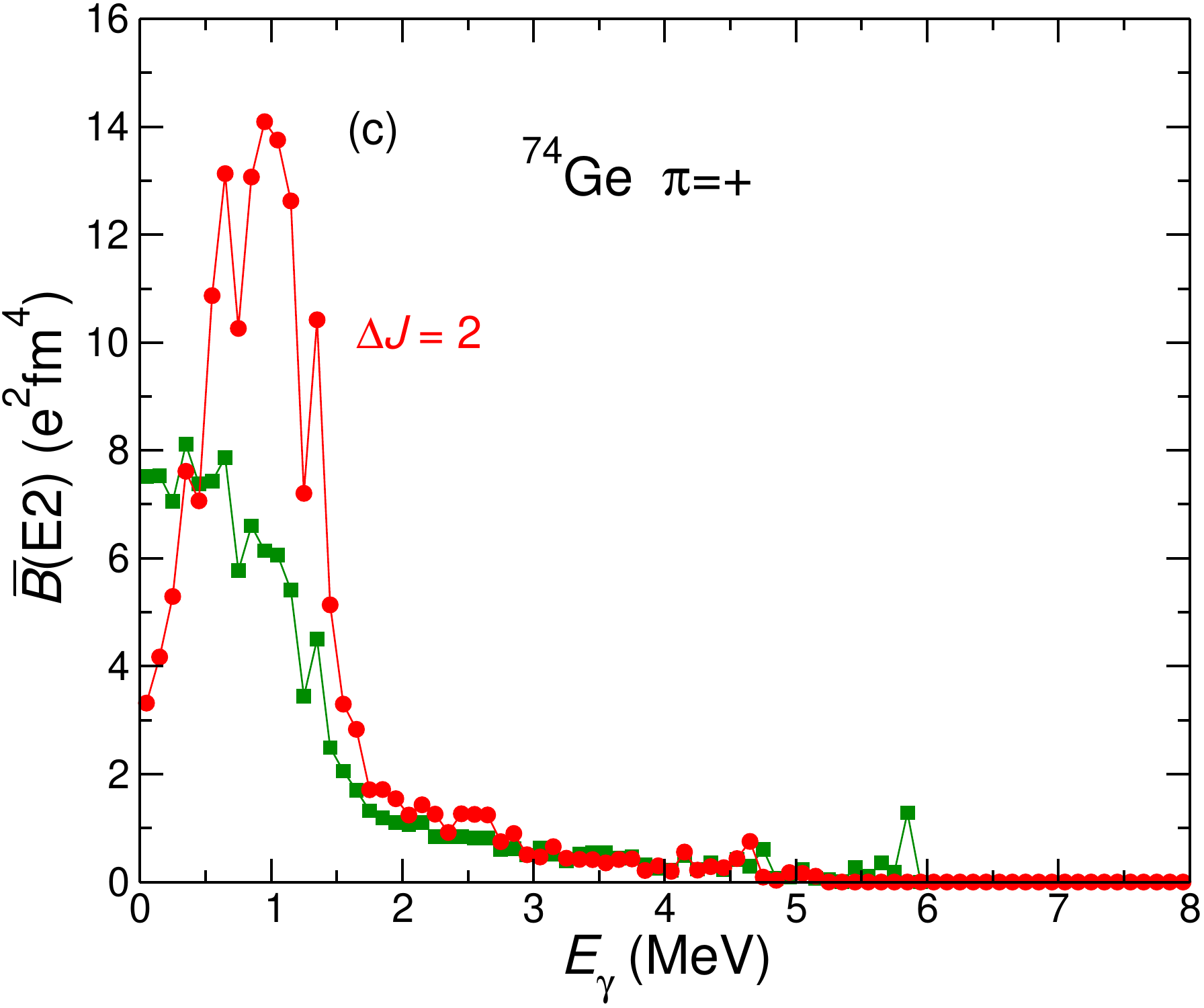,width=5.6cm}  
\caption{\label{fig:A74} As Fig.~\ref{fig:A64}, but for $^{74}$Ge.}
\end{figure*}

\begin{figure*}
\epsfig{file=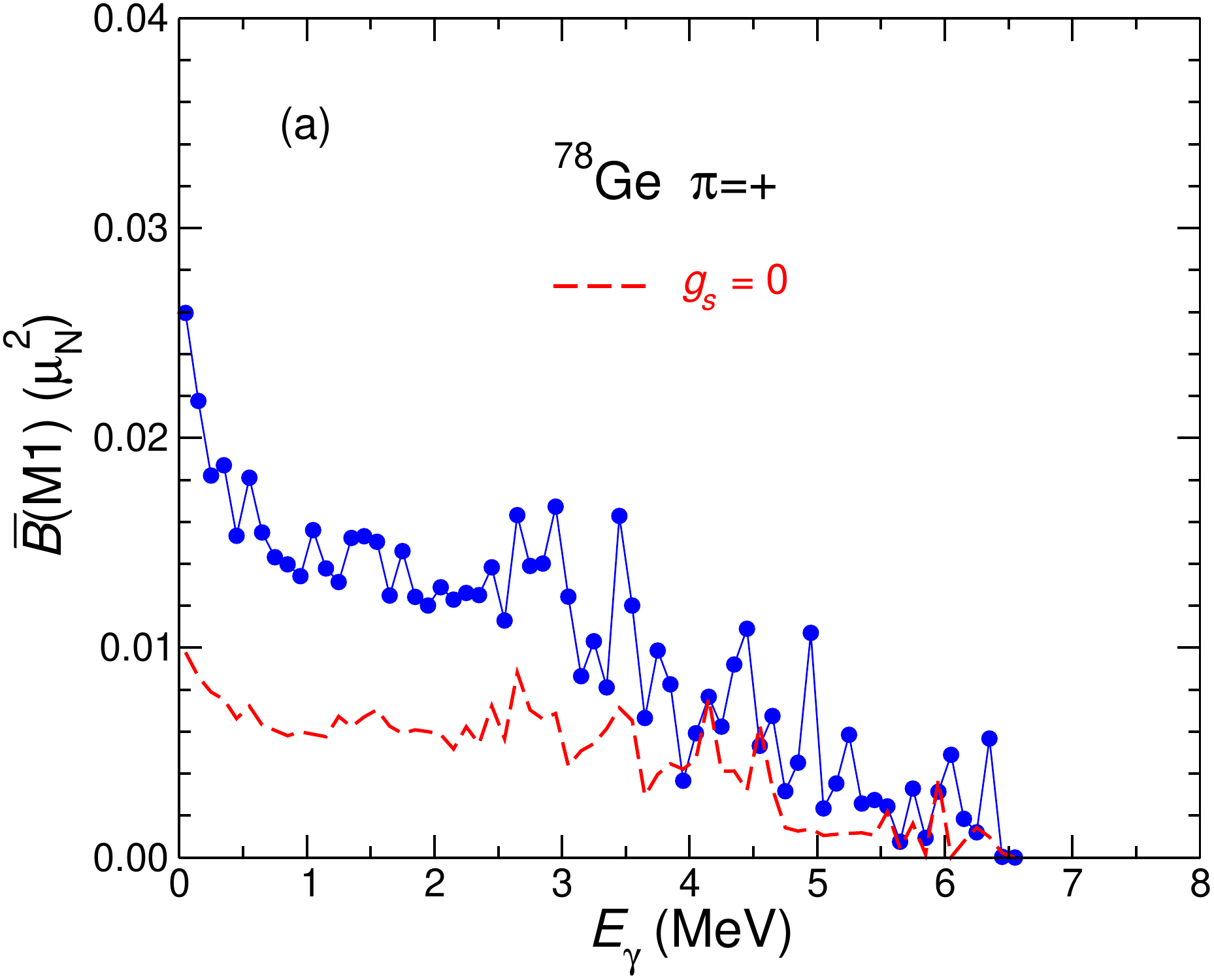,width=5.8cm}
\epsfig{file=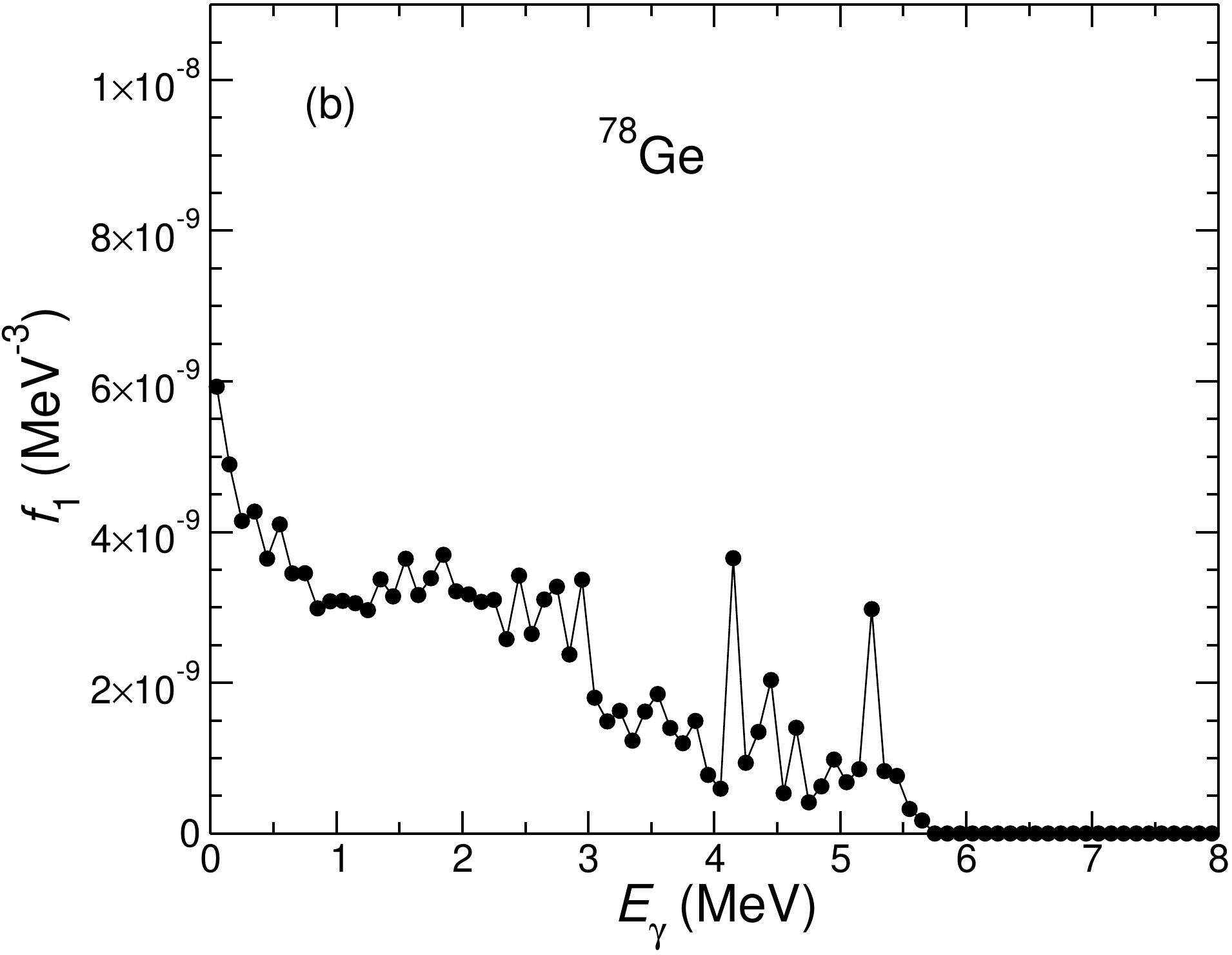,width=5.9cm}
\epsfig{file=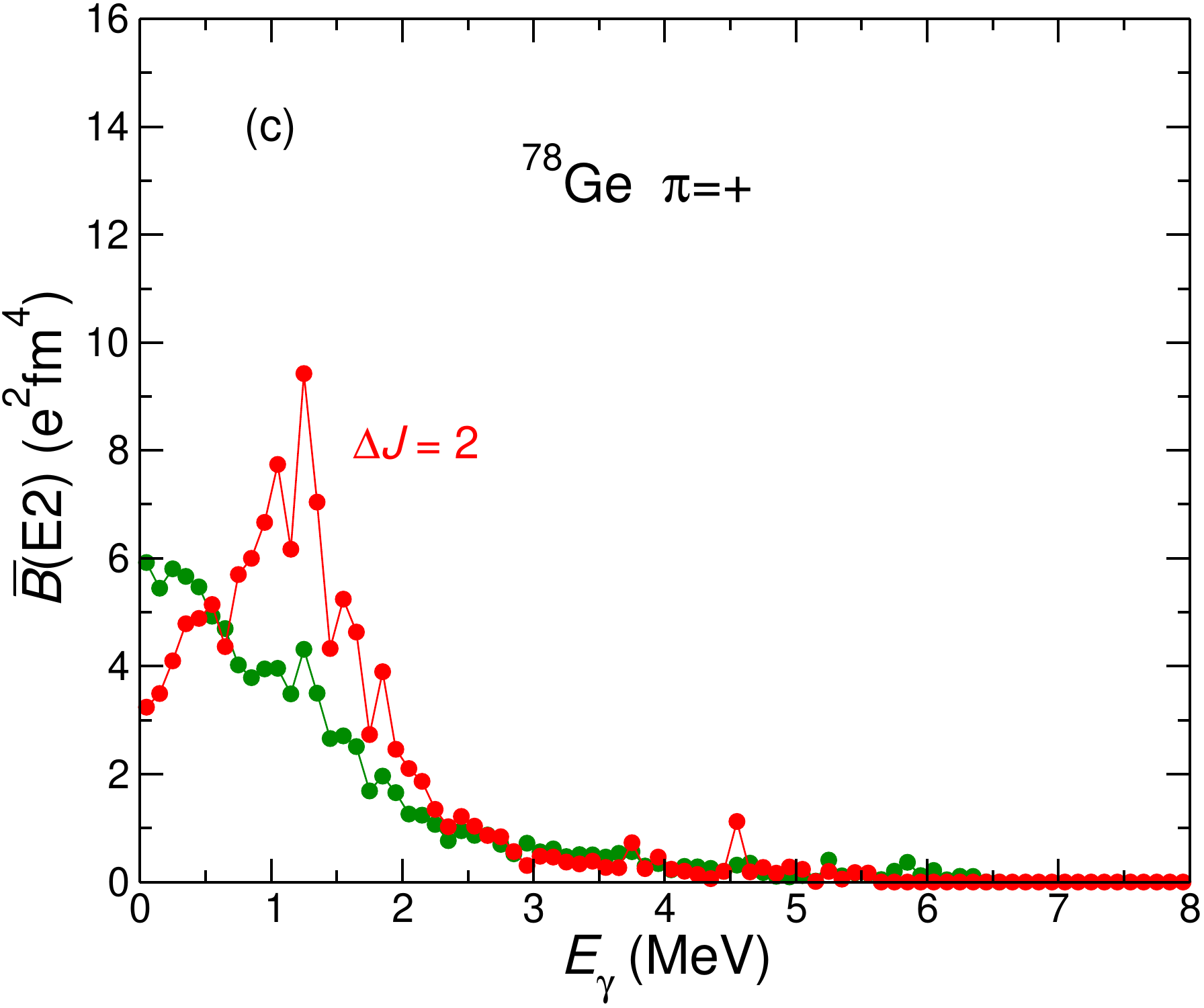,width=5.6cm}  
\caption{\label{fig:A78} As Fig.~\ref{fig:A64}, but for $^{78}$Ge.}
\end{figure*}

\begin{figure*}
\epsfig{file=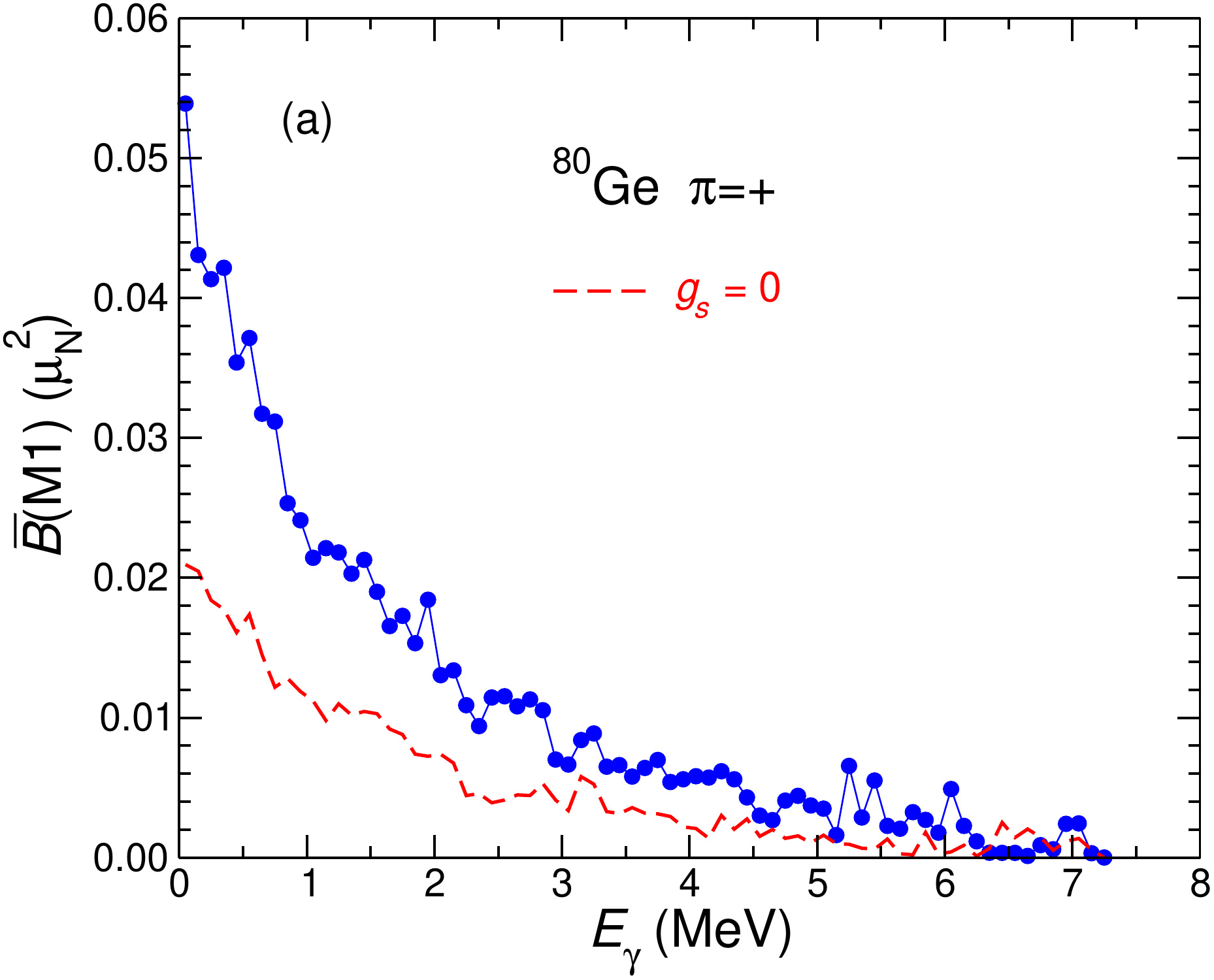,width=5.8cm}
\epsfig{file=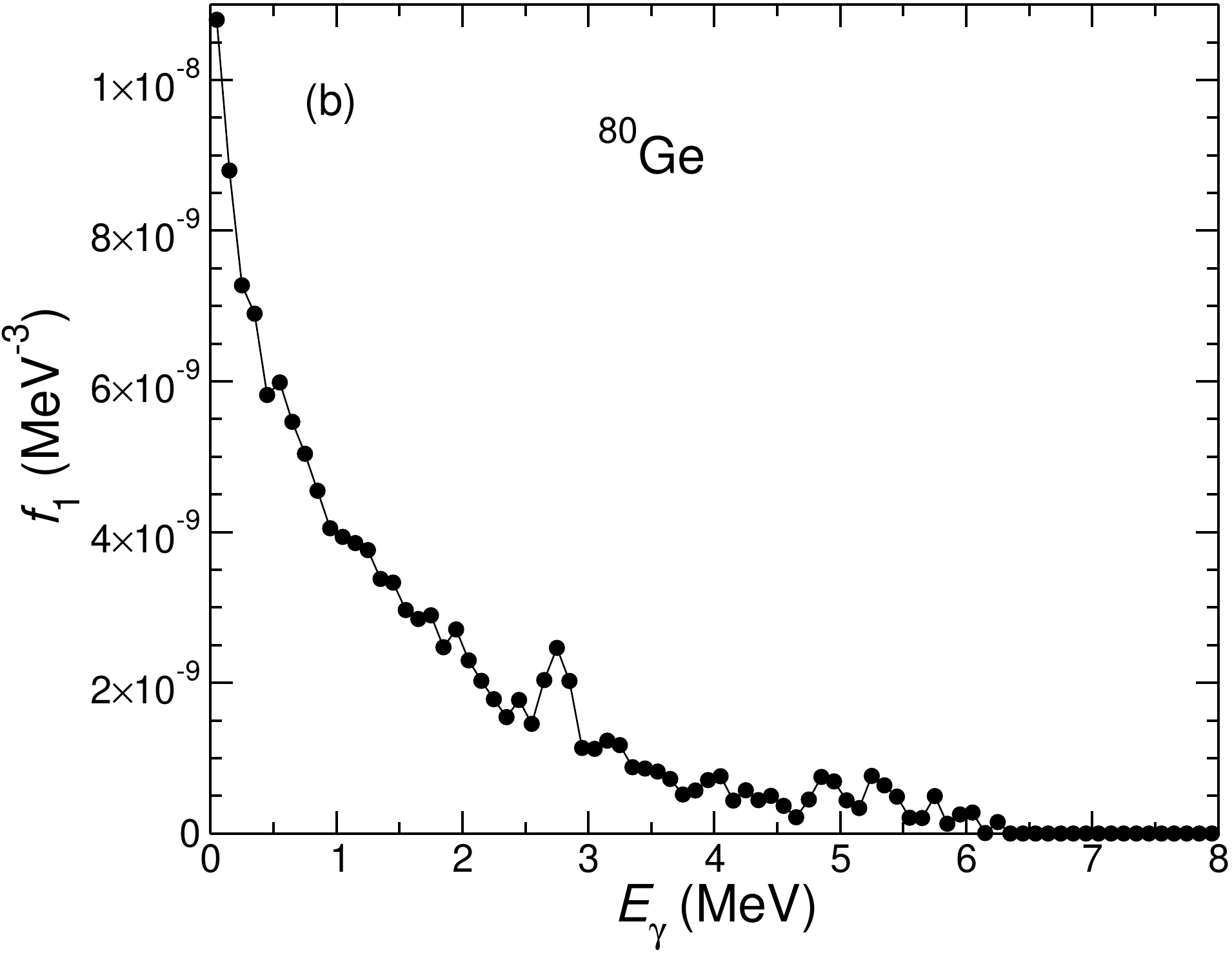,width=5.9cm}
\epsfig{file=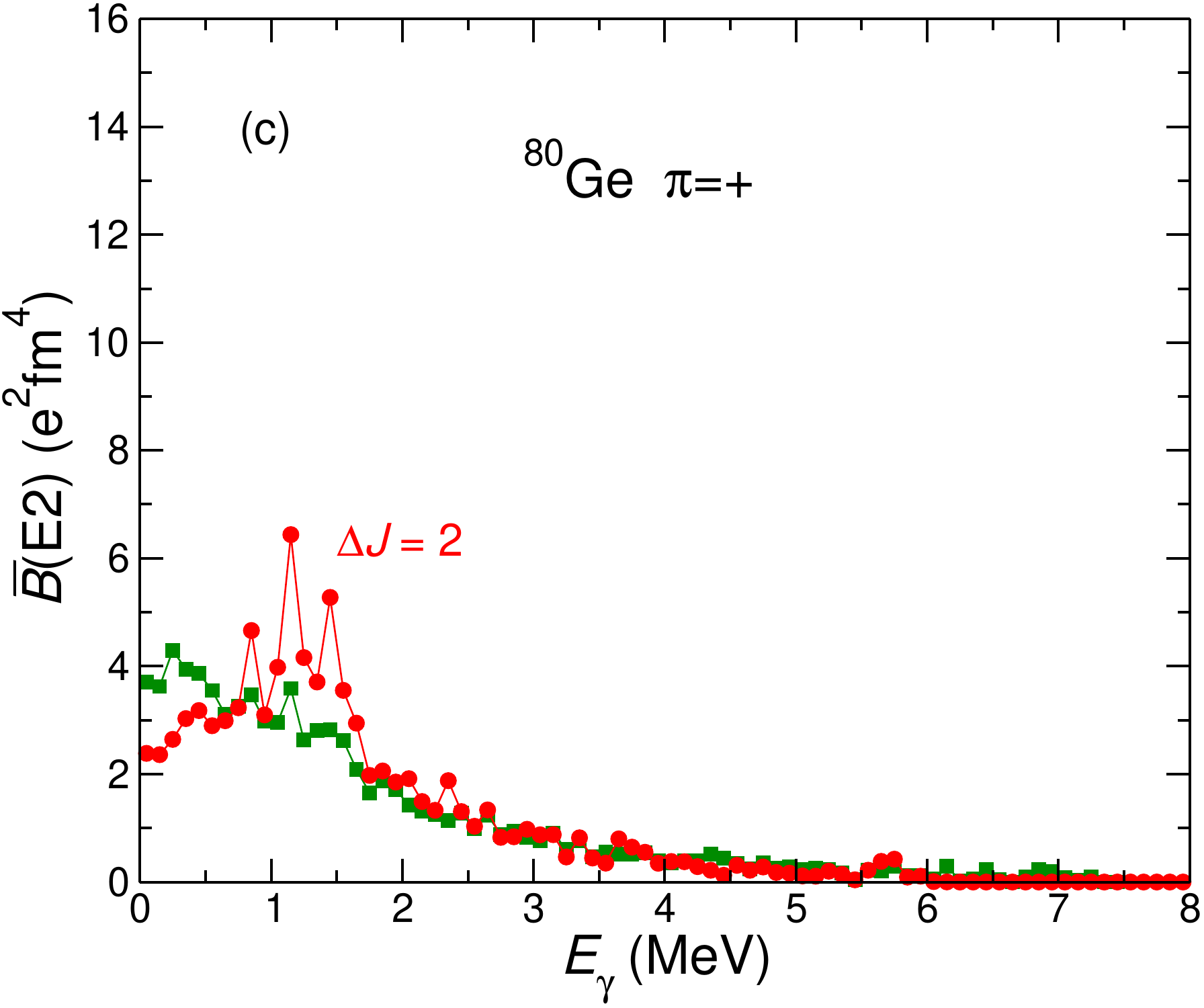,width=5.6cm}  
\caption{\label{fig:A80} As Fig.~\ref{fig:A64}, but for $^{80}$Ge. Note the
  different vertical scale in (a) compared with the corresponding graphs for
  the other isotopes.} 
\end{figure*}

\section{Results for the strength functions}
\label{sec:res}

The average $B(M1)$ and $B(E2)$ values for positive parity states and the
$M1$ strength functions including both parities are shown for all considered
Ge isotopes in Figs.~\ref{fig:A64},\ref{fig:A66},\ref{fig:A70},\ref{fig:A74},\ref{fig:A78},\ref{fig:A80}.
The $N = Z = 32$ nucleus $^{64}$Ge shows a fluctuating, but on average flat
distribution of the $B(M1)$ strength as a function of $E_\gamma$ with an even
slight decrease toward $E_\gamma = 0$, which is similar to predictions for
the $N = Z$ nuclei $^{48}$Ca ~\cite{sie17a} and $^{108}$Xe \cite{sie18}. This
seems to point to a more general feature of the low-energy $M1$ strength in
$N = Z$ nuclei. It is suggested below that
 isospin conservation quenches the LEMAR spike. 
In the $N = Z + 2$ nuclide $^{66}_{32}$Ge$_{34}$, a gradual enhancement of the
$M1$ strength toward $E_\gamma = 0$ is seen.  The behavior
resembles the one in $^{60}_{26}$Fe$_{34}$ \cite{sch17}, but is less pronounced.
Both these nuclei are localized near the bottom of the neutron $(fpg)$ shell.
For $^{70}_{32}$Ge$_{38}$, the bimodal structure of a LEMAR peak at
$E_\gamma = 0$ and a broad SR peak around  $E_\gamma = 3$ MeV appears. A similar
bimodal distribution is seen in $^{74}_{32}$Ge$_{42}$. This bimodal strength
distribution is characteristic for nuclei located well in the open shell, as
for example also in $^{64,68}_{~~~26}$Fe$_{38,42}$ \cite{sch17} and in nuclides
with $A >$ 100 \cite{sie18}.
The SR peak becomes weak in $^{78}_{32}$Ge$_{46}$ and disappears in
$^{80}_{32}$Ge$_{48}$, when approaching the top of the  neutron shell.
A similar suppression of the SR peak toward the next higher neutron shell was
also found in $N \approx 80$ nuclei \cite{sie18}.
The present calculations find a maximum of the SR strength in the middle of
the neutron shell that correlates with the clear maximum of the experimental
$B(E2, 2^+_1\rightarrow 0^+_1)$ values. Unlike the experiment, the calculated 
$B(E2,2^+_1\rightarrow 0^+_1)$ values have a very shallow mid-shell maximum,
as was also obtained in the calculations  for the Fe isotopes~\cite{sch17}. 

However, a different behavior is seen for the average
$B(E2,J \rightarrow J - 2)$ values shown in the panels (c) of 
Figs.~\ref{fig:A64},\ref{fig:A66},\ref{fig:A70},\ref{fig:A74},\ref{fig:A78},\ref{fig:A80}. 
Here, a peak around 1 MeV develops toward the middle of the
shell ($^{70,74}$Ge). This peak indicates enhanced collectivity in the
$\Delta J = 2$ sequences and clearly correlates with the maximum of the SR
strength in the middle of the shell. It can be interpreted as the appearance
of damped rotational transitions (see e.g. Ref.~\cite{mat97}) as a consequence
of building-up quadrupole collectivity. The calculations include states up
angular momentum 10 $\hbar$ with equal weight. The average transition energy
of 1 MeV and the average angular momentum of $5~\hbar$ correspond to a moment
of inertia of about 10 $\hbar^2$/MeV, which is somewhat smaller than the
rigid-body value of 14 $\hbar^2$/MeV for $A = 70$. In the same way, from
Fig.~4 of Ref.~\cite{sch17} one derives a moment of inertia of
8 $\hbar^2$/MeV for $^{68}_{26}$Fe$_{42}$. In Ref.~\cite{sim16}, the increased
SR strength of the $\gamma$-SF of $^{151,153}$Sm could be reproduced by
replacing the ground state moment of inertia by the rigid-body value in the
phenomenological expression of Ref. \cite{end05}, which was developed for the
excitation of the SR from the ground state.  

We calulated the total $B(M1)$ strengths in certain energy ranges by a
numerical integration of the $M1$ strength functions:

\begin{equation}
\label{eq:Btot}
B(M1)_{\rm tot} =
9/(16\pi)~(\hbar c)^3~\sum f_{M1}(E_\gamma) \Delta E_\gamma.
\end{equation}

The results for the LEMAR region ($E_\gamma < 2$ MeV), the SR region
(2 MeV $\leq E_\gamma < 5$ MeV) and their sums are compiled in
Table~\ref{tab:Btot}. As visualized in the panels (b) of
Figs.~\ref{fig:A64},\ref{fig:A66},\ref{fig:A70},\ref{fig:A74},\ref{fig:A78},\ref{fig:A80}, 
one also quantitatively observes a shift of strength to the SR
region when going into the open shell ($^{70,74}_{~~~32}$Ge$_{38,42}$) and a
shift back to the LEMAR region when approaching the $N$ = 50 shell closure
($^{78,80}_{~~~32}$Ge$_{46,48}$), while the sum of the two remains roughly
constant. Only the $N = Z$ nuclide $^{64}$Ge does not fit the systematics for
reasons discussed below. 
In the calculations of  Ref.~\cite{sch17}, a similar redistribution of the $M1$
strength was found for the isotopes $^{60,64.68}_{~~~~~~26}$Fe$_{34,38,42}$ when
moving into the open shell by adding neutrons. However, the integrated strength
up to 5 MeV, which is the sum of the LEMAR and SR strength, is about
1 $\mu_N^2$ for the Ge isotopes and  about 10 $\mu_N^2$ for the Fe isotopes.
The authors of Ref.~\cite{sch13L} suggested that the low-energy $M1$ radiation
is generated by the reorientation of the valence nucleons on high-$j$ orbitals.
This mechanism is particular efficient if protons are hole-like and neutrons
are particle-like (or vice versa). Then the transverse magnetic moments add up,
which generates strong $M1$ radiation. An analogous mechanism generates the
"shears bands" manifesting "magnetic rotation" \cite{MagRot}. In the case of
the Fe isotopes one has active $1f_{7/2}$ proton holes, which favorably combine
with the active $1g_{9/2}$ neutrons. In the case of the Ge isotopes the active
$1g_{9/2}$ neutrons combine with the $1f_{5/2}$ protons, which have a small
magnetic moment, and  $1g_{9/2}$ protons, which have a magnetic moment with the
opposite sign. The factor of 10 in the integrated low-energy $M1$ strength
reflects the different valence proton configurations of the Fe and Ge isotopes. 
In case of the $Z > 50$, $N \geq 80$ nuclides the integrated strength up to
4 MeV is 0.5 - 1 $\mu_N^2$ (see Fig.~3 of Ref.~\cite{sie18}). The small number
is expected because active protons and neutrons are particle-like and do not
occupy the high-$j$ orbitals.    

Also shown in Table~\ref{tab:Btot} are the integrated $B(E2)$ strengths up to
$E_\gamma$ = 5 MeV, which were determined analogously to the integrated $B(M1)$
strengths. The values are maximal at the mid-shell nuclei $^{70,74}$Ge in
accordance with the SR strengths, which proves the correlation of SR strength
and collectivity.

\begin{figure}
\epsfig{file=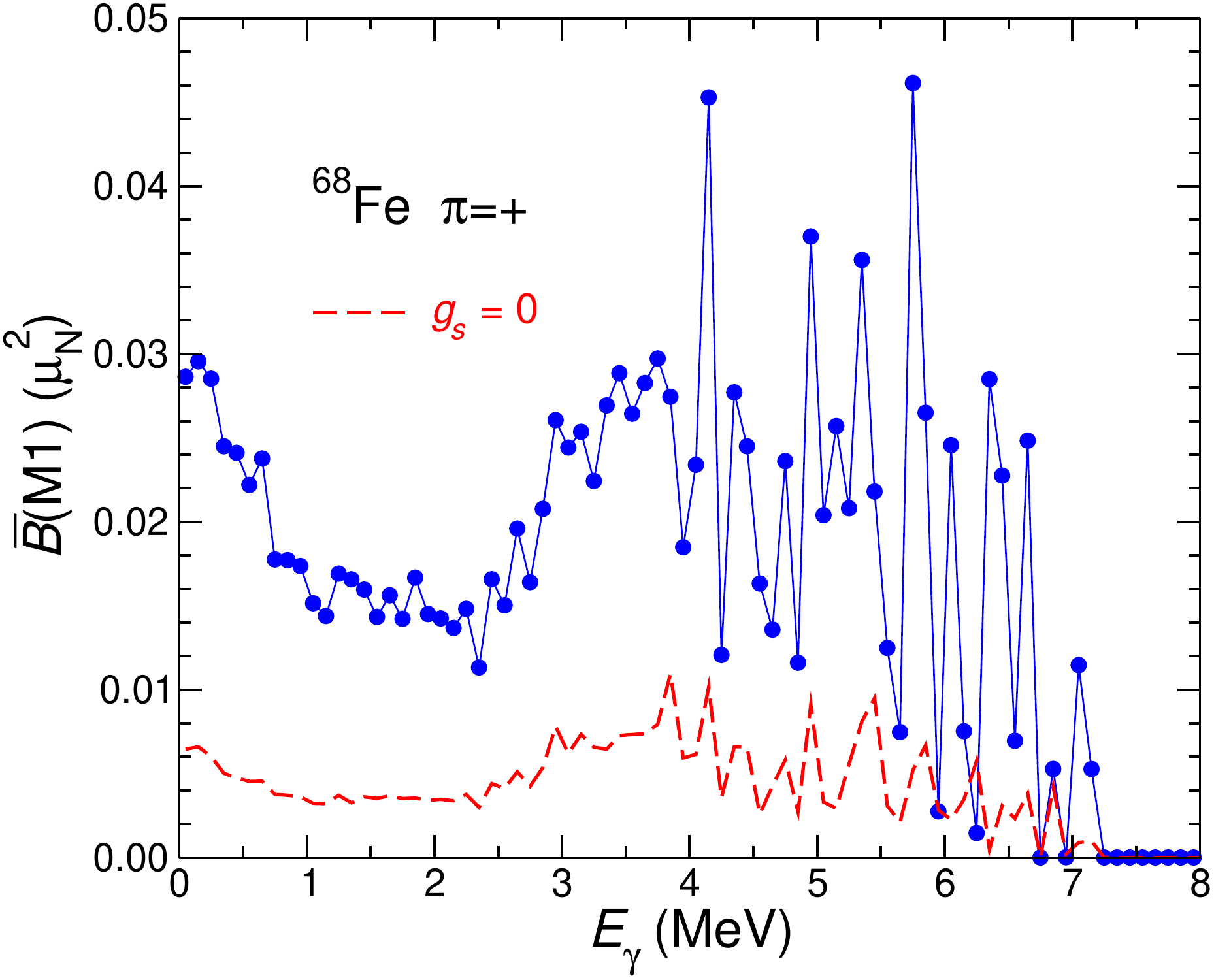,width=5.8cm}  
\caption{\label{fig:68Fe} As Fig.~\ref{fig:A64}(a), but for
  $^{68}$Fe. The corresponding shell-model calculations are described in
  Ref.~\cite{sch17}. }
\end{figure}

\begin{figure}
\epsfig{file=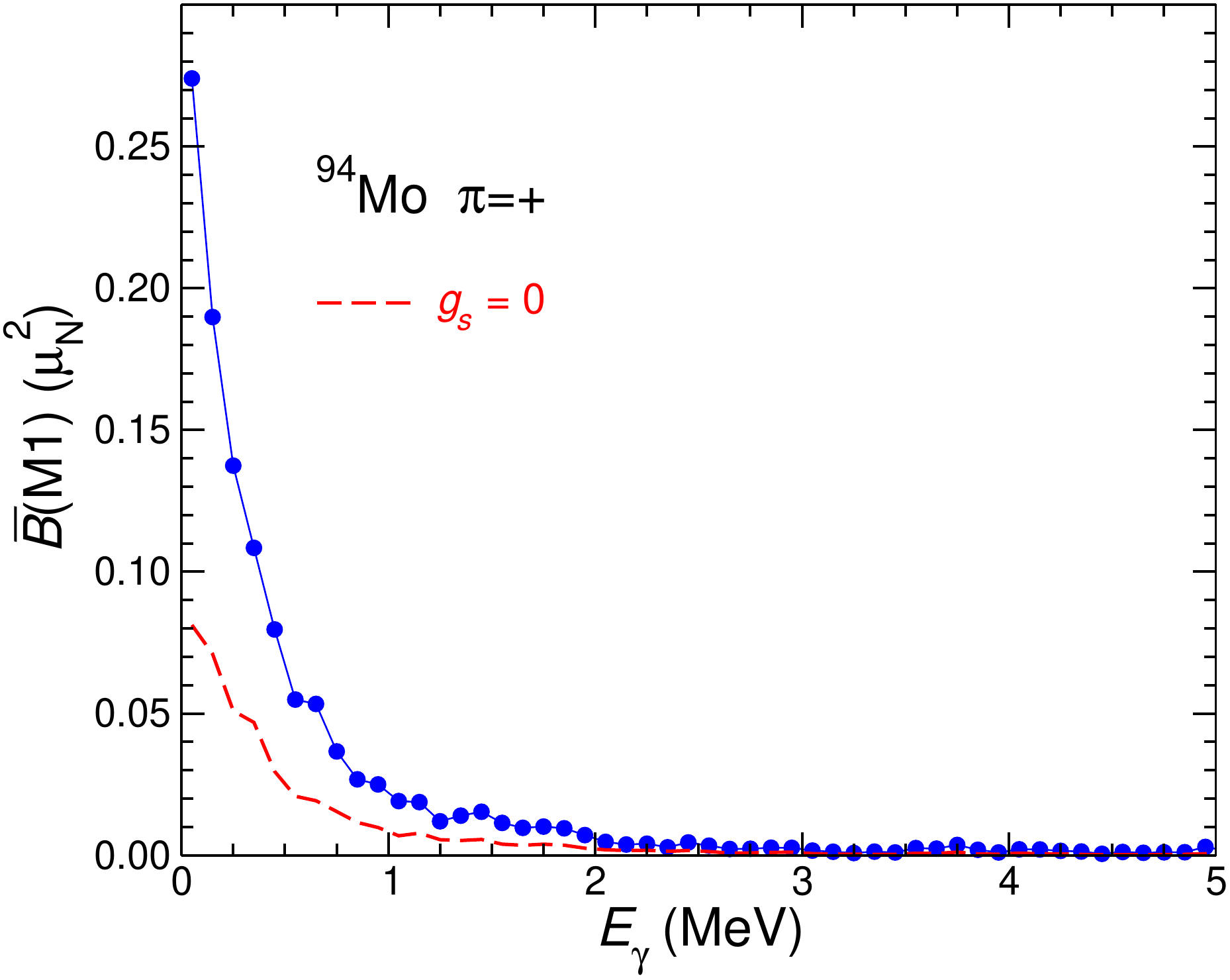,width=5.8cm}  
\caption{\label{fig:94Mo} As Fig.~\ref{fig:A64}(a), but for
  $^{94}$Mo. The corresponding shell-model calculations are described in
  Ref.~\cite{sch13L}.}
\end{figure}

The excitation of the SR from the ground state, which appears as a bunch of
$1^+$ states around 3 MeV, has been extensively studied and reviewed in
Ref.~\cite{hey10}. Therein, the SR is considered as being dominated by
exciting the orbital angular momentum of the protons.
In contrast, our calculations for the Ge isotopes show a reduction of the
low-energy strength by a factor of about two when the spin part of the
magnetic dipole operator is set equal to zero, which is illustrated in the
panels (a) of Figs.~\ref{fig:A64},\ref{fig:A66},\ref{fig:A70},\ref{fig:A74},
\ref{fig:A78},\ref{fig:A80}. An equal reduction appeared in our earlier
calculations for the Mo and Fe isotopes as displayed in Figs.~\ref{fig:68Fe}
and \ref{fig:94Mo}. The strong dependence of the $M1$ strength on the spin part
indicates that the reorientation of high-$j$ orbitals such as $1g_{9/2}$ and
$1h_{11/2}$ must play a central role in generating the strong $M1$ radiation
(see Ref.~\cite{sch17}). In the calculations for $^{130}$Te in
Ref.~\cite{sie18}, the SR peak is similarly quenched when the spin part of the
$M1$ operator is set equal to zero whereas the LEMAR spike remains unchanged
(see Figs.~2c and 2d therein).

The SR strength in the medium and heavy nuclei has been found proportional to
the square of the quadrupole deformation, which is maximal in the middle of the
open shell \cite{hey10}. As discussed in Ref.~\cite{sim16}, the strength
of the SR in the $\gamma$-SF seems to be related to the nuclear ground-state
deformation in a similar way, though being a factor of three larger. The
calculations in Ref.~\cite{sie18} show the built-up of a SR around 4 MeV when
moving into the open shell by adding neutrons in the case of the Te isotopes
and when adding protons in the case of the $N = 80$ isotones, where the SR
strength increases with the $B(E2, 2^+_1\rightarrow 0^+_1)$ values (compare
Figs.~1 and 3 in Ref.~\cite{sie18} and the experimental values for the studied
nuclides given in Ref.~\cite{ram01}).  The same correlation between the
increase of experimental $B(E2,2^+_1\rightarrow 0^+_1)$ and of the increase of
the SR strength was found for the Fe isotopes in the calculations in
Ref.~\cite{sch17}. 

The integrated strengths up to a transition energy of 5 MeV in
Table~\ref{tab:Btot} are about 1 $\mu_N^2$. They are much larger than the sums
of the strengths of transitions from $1^+$ states down to the ground state,
which are also given in Table~\ref{tab:Btot}. As seen in Table~II of
Ref.~\cite{sch17}, the integrated strengths up to a transition energy of 5 MeV
in $^{60,64,68}$Fe are about 10 $\mu_N^2$ to be compared with 0.33, 0.55, 0.58
$\mu_N^2$ for the respective sums of all transitions from $1^+$ states down to
the ground state. In Ref.~\cite{sim19}, the experimental integrated $M1$
strengths up to 5 MeV for transitions in the quasicontinuum of
$^{147, 149,151,153}$Sm were found to be  about 8 $\mu_N^2$, which has to be
compared with summed strengths of 0.16, 0.32, 0.80, 0.81 $\mu_N^2$ from
$1^+$ states to the ground states of $^{148, 150,152,154}$Sm, respectively
\cite{end05}. In Ref.~\cite{sch17}, the enhancement was attributed to the
quenching of the pair correlations with increasing excitation energy, i.e. the
thermal quenching of pairing. 

According to the collective model of Ref.~\cite{end05}, the $M1$ strength of
the SR on the ground state scales $\propto A\delta^2$, where $\delta$ is the
deformation parameter and $A$ the mass number. The sums
$\sum B(M1, 1^+ \rightarrow 0^+_1)$ of transition down to the ground states
of the Ge, Fe,Te and Sm isotopes roughly follow the scaling 
(cf.  Table~\ref{tab:Btot} and Refs.~\cite{sch17,sie18,sim19}). The simple
collective behavior seems to be caused by the pair correlations. The
integrated $M1$ strengths for the transitions in the quasicontinuum do not
obey it.  Once the pair correlations  are quenched, the bimodal LEAMAR-SR 
structure appears in the strength function, the total strength of which depends
strongly on the individual magnetic properties of the valence nucleons. This
is in analogy to the moments of inertia. At low spin, when the  the pair
correlations are strong, the moments of inertia behave in a systematic manner,
being  $\propto A^{5/3}\delta^2$. At high spin, when the Coriolis force
overcomes the pair correlation, the individuality of the valence nucleons
comes to light.

The experimental summed strengths for the transitions from $1^+$ states to the
ground state given in Table~\ref{tab:Btot} are smaller than the calculated
ones. However, the experimental values represent only a lower limit. In the
calculations, a large number of weak transitions contributes to the summed
strengths, whereas experiments as the ones in Ref.~\cite{jun95} detect only
the strongest transitions. It has been demonstrated that a large number of
weak transitions, which are hidden in a quasicontinuum, may substantially
enlarge the $M1$ strength function, as seen for example in Ref.~\cite{mas14}.

The $M1$ operator has approximately isospin $T = 1$ character. In $N = Z$
nuclei the low-lying states have $T = 0$. The $T = 1$ states lie substantially
higher. In case of $^{64}$Ge, the experimental energy of the lowest $T = 1$
state is 6.2 MeV. Thus, the sum
$\sum B(M1, 1^+ \rightarrow 0^+_1)$ = 0.001 $\mu_N^2$ in Table~\ref{tab:Btot}
includes only transitions between $T = 0$ states, which are isospin forbidden.
The very small value for  $^{64}$Ge  reflects that isospin conservation nearly
quenches $M1$ transitions between the $T = 0$ states. For $N = Z + 2$ nuclei
the low-lying states have $T=1$. Transitions between  $T = 1$ states are
allowed, which results in $\sum B(M1, 1^+ \rightarrow 0^+_1) = 0.25 \mu_N^2$
for $^{66}$Ge. One expects that the same mechanism works for higher excitation
energies. Transitions between the $T = 0$ states are nearly forbidden.
The $T = 1$ states lie on the average substantially above the $T = 0$ states
that are connected by the $M1$ operator, which prevents transition energies
close to zero. For $N > Z$ nuclides, the transitions between the states with
the same isospin $T > 0$ are allowed and the LEMAR spike appears. The author of
Ref.~\cite{sie18} suggested an alternative explanation:
$N = Z$ nuclei have a particular large deformation that moves $M1$ strength
from the LEMAR spike to the SR, which results in a flat distribution. At
variance, the $N$ dependences of the $E2$ strength in
Figs.~\ref{fig:A64} to \ref{fig:A80} and Table~\ref{tab:Btot}
indicate little $E2$ collectivity for $^{64}$Ge.

\begin{table}
  \caption{\label{tab:Btot}Summed $B(M1)$ strengths in ranges of transition
    energy (LEMAR: $E_\gamma < 2$ MeV, Scissors: 2 MeV $\leq E_\gamma < 5$ MeV,
    $\sum$: the sum of the two), summed $B(E2)$ strengths for
    $E_\gamma <$ 5 MeV, and summed strengths of all discrete transitions from
    $1^+$ states to the ground states in $^{64,66,70,74,78,80}$Ge.}
\begin{ruledtabular}
\begin{tabular}{ccccccc} 
  & \multicolumn{3}{c}{$B(M1)_{\rm tot}$\footnotemark[1]} &
                  $B(E2)_{\rm tot}$\footnotemark[2] &
\multicolumn{2}{c}{$\sum B(M1,1^+ \rightarrow 0^+_1)$\footnotemark[3]} \\
  & \multicolumn{3}{c}{$(\mu_N^2)$} & ($e^2$fm$^4$) &
                                     \multicolumn{2}{c}{$(\mu_N^2)$}        \\
\cline{2-4} \cline{6-7}                                      \\
           & LEMAR & SR & $\sum$ &   & EXP\footnotemark[4] & CALC           \\
\hline
$^{64}$Ge$_{32}$ & 0.30 & 0.54 & 0.84 & 155 &         & 0.001 \\
$^{66}$Ge$_{34}$ & 0.35 & 0.35 & 0.70 & 185 &         & 0.25  \\
$^{70}$Ge$_{38}$ & 0.54 & 0.62 & 1.16 & 219 & 0.04(1) & 0.49  \\
$^{74}$Ge$_{42}$ & 0.44 & 0.50 & 0.94 & 241 & 0.30(3) & 0.57  \\
$^{78}$Ge$_{46}$ & 0.63 & 0.49 & 1.12 & 181 &         & 0.48  \\
$^{80}$Ge$_{48}$ & 0.84 & 0.28 & 1.12 & 123 &         & 0.31  \\
\end{tabular}
\end{ruledtabular}
\footnotetext[1]{Integrated $M1$ strength calculated according to 
  Eq.~(\ref{eq:Btot}).}
\footnotetext[2]{Integrated $E2$ strength calculated for positive-parity states
   in analogy to Eqs.~(\ref{eq:f1M1}) and (\ref{eq:Btot}).}
\footnotetext[3]{Summed $M1$ strength of transitions from the
   $1^+$ states below 5 MeV to the ground state.}
\footnotetext[4]{Value taken from Ref.~\cite{jun95}.}
\end{table}

\section{Summary}
\label{sec:sum}

Shell-model calculations were performed for the series of germanium isotopes
with neutron numbers from $N$ = 32 to $N$ = 48. Average $B(M1)$ and $B(E2)$
strengths were determined from a large number of transitions linking states of
spins from 0 to 10. The average $B(M1)$ strengths and the associated $M1$
strength functions are strongly enhanced near zero transition energy, which is
the LEMAR spike observed before. The LEMAR spike develops with increasing
neutron number and  is strongest at $N = 80$. It is suppressed at $N = Z$,
which is attributed to isospin conservation. In the mid-shell nuclei, a bump
around 3.5 MeV  appears, which is interpreted as the scissors resonance. The
strength of the SR correlates with the quadrupole collectivity, as reflected by
the experimental $B(E2,2^+_1\rightarrow 0^+)_1$ values and the integrated
average $E2$ strength of quasicontinuum transitions. 
The sum of the LEMAR and SR strengths depends only weakly on the neutron number.
These characteristics are consistent with those found for the series of iron
isotopes and with the experimental observation of LEMAR and SR strengths in
samarium isotopes as well. They exhibit the important role of high-$j$
orbitals, such as $1g_{9/2}$ and $1h_{11/2}$, for the evolution of the low-lying
modes. Spin and orbital contributions to the $M1$ strength appear nearly equal
at low energy in most isotopes, while there are stronger orbital contributions
above 4 MeV of transition energy in the mid-shell isotopes $^{70,74}$Ge.
The present systematic analysis of low-lying $M1$ strength in a relatively long
isotopic series demonstrates that the correlated appearance of the two $M1$ 
modes is a phenomenon that occurs across various mass regions.

\section{Acknowlegdments}

We thank B. A. Brown for his support in using the code NuShellX@MSU.
The allocation of computing time through the Centers for High-Performance
Computing of Technische Universit\"at Dresden and of Helmholtz-Zentrum
Dresden-Rossendorf are gratefully acknowledged.  S. F. acknowledges support
by the DOE Grant DEFG02-95ER4093.

\clearpage
 \onecolumngrid
 \begin{appendix}
\newpage
 
 \section{Yrast properties of the remaining isotopes}
\begin{figure}[h]
\epsfig{file=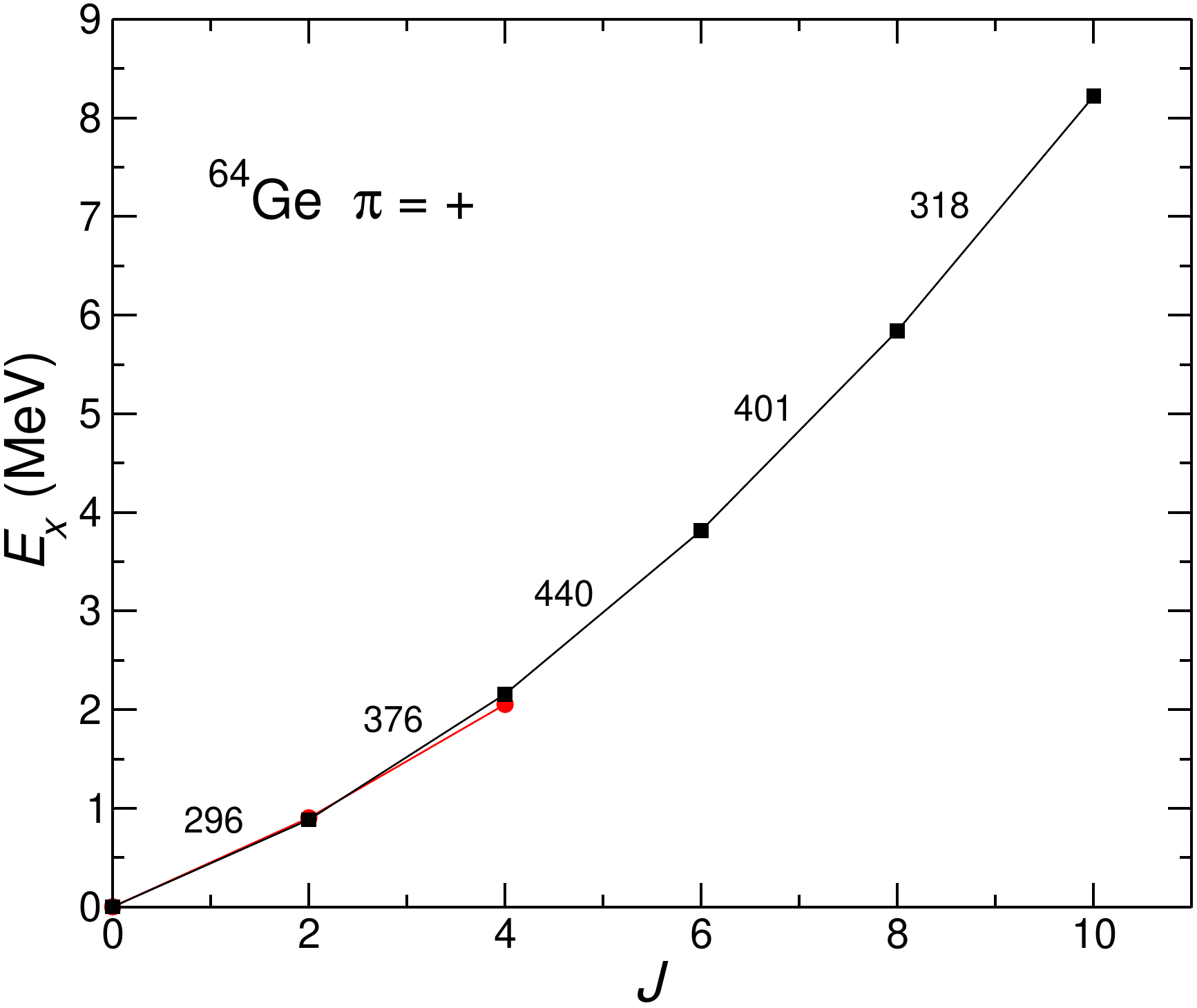,width=5.8cm}  
\epsfig{file=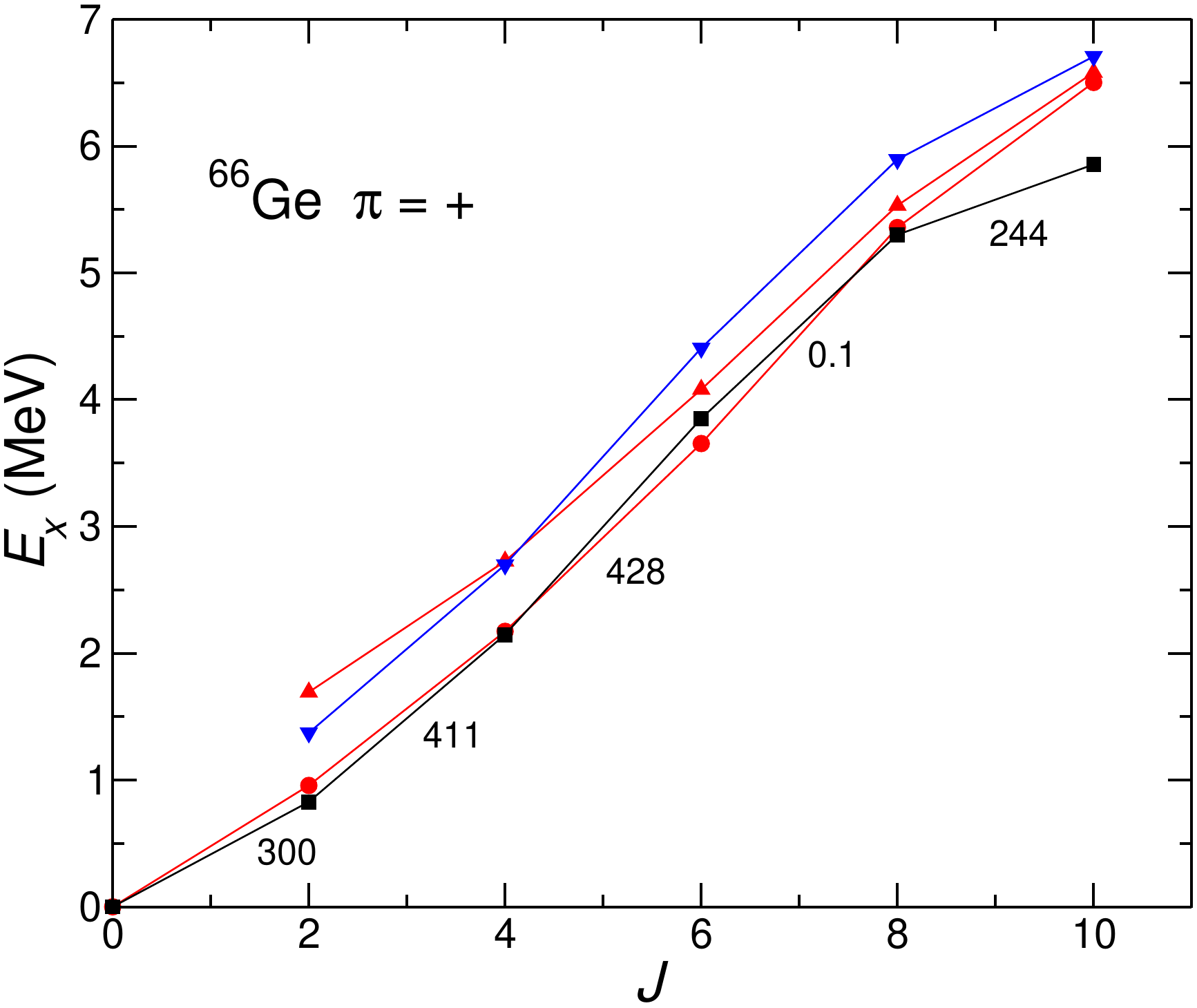,width=5.8cm}  
\epsfig{file=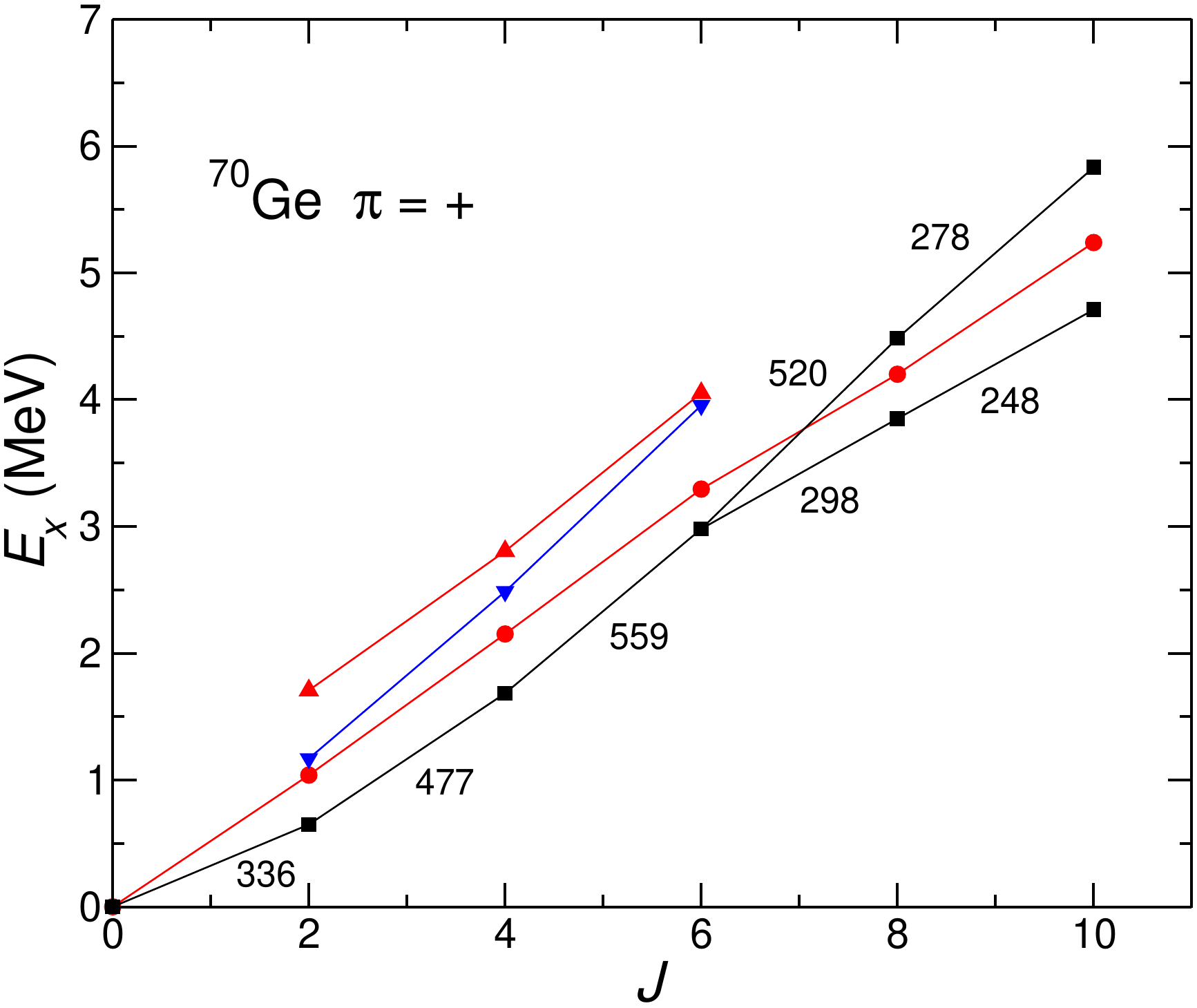,width=5.8cm}  
\caption{\label{fig:64,66,70GeJE} Excitation energies versus spin of
  experimental (red circles) and calculated (black squares) yrast states in
  $^{64,66,70}$Ge, and of experimental (red triangles up) and calculated
  (blue triangles down) states built on the second $2^+$ states in $^{66,70}$Ge.
  The lines represent the linking $E2$ transitions. The numbers at the lines are
  calculated $B(E2)$ values in e$^2$fm$^4$. The experimental data were taken
  from Refs.~\cite{far03,ste03,muk01}, respectively.}
\end{figure}

\begin{figure}[h]
\epsfig{file=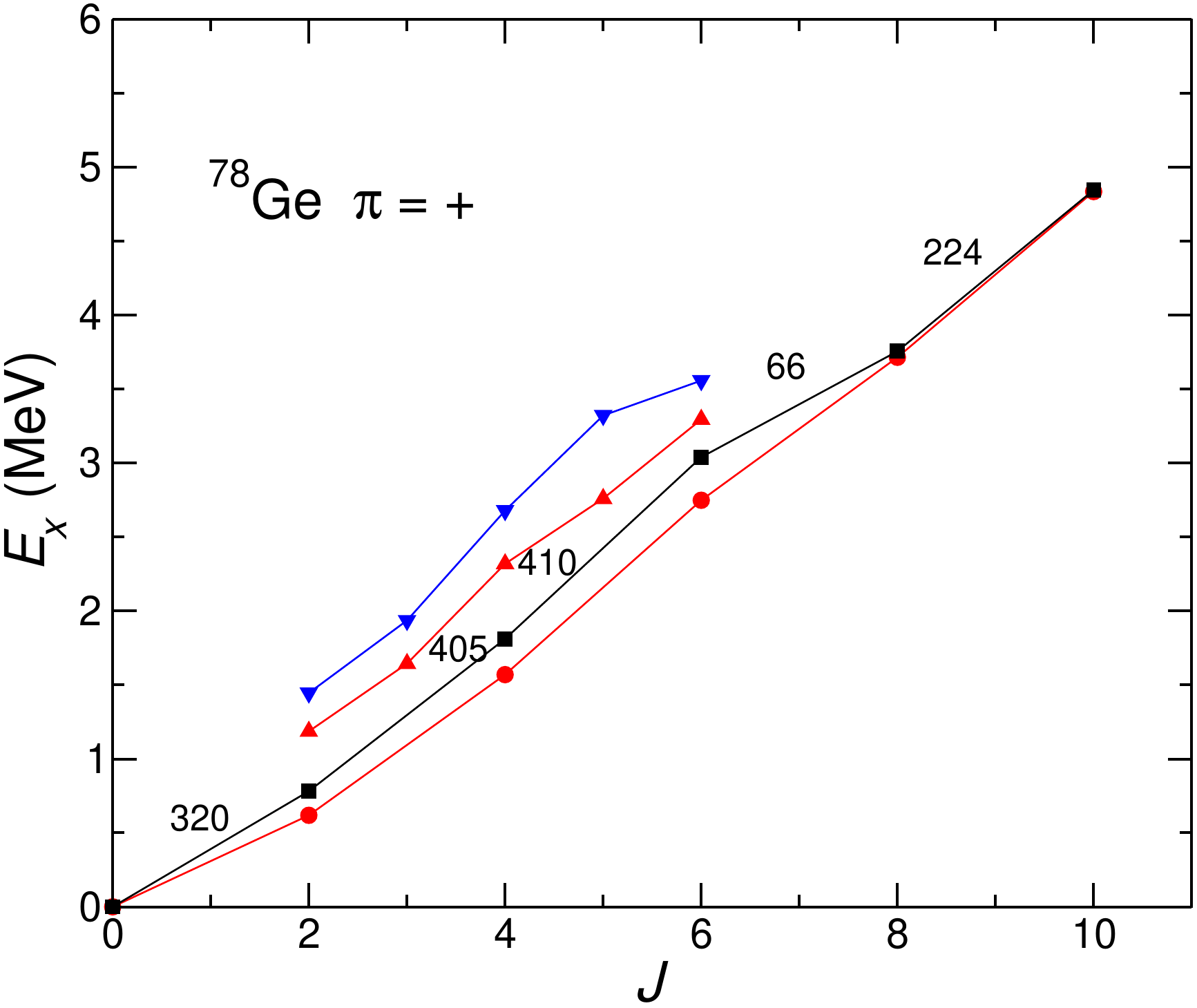,width=5.8cm}  
\epsfig{file=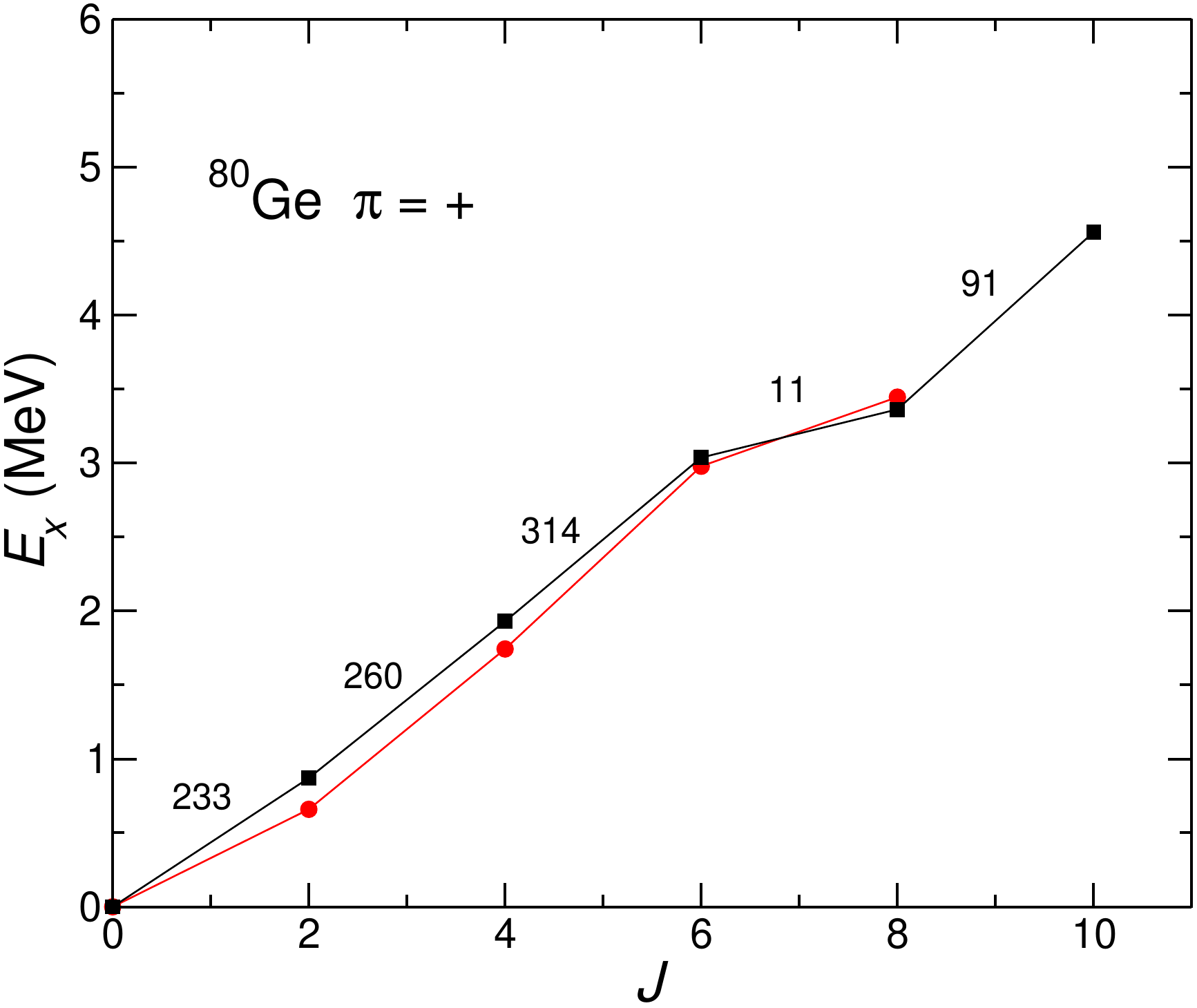,width=5.8cm}
\caption{\label{fig:78,80GeJE} Excitation energies versus spin of experimental
  (red circles) and calculated (black squares) yrast states in $^{78,80}$Ge,
  and of experimental (red triangles up) and calculated (blue triangles down)
  states built on the second $2^+$ state in $^{78}$Ge. The lines represent the
  linking $E2$ transitions. The numbers at the lines are calculated $B(E2)$
  values in e$^2$fm$^4$. The experimental data were taken from
  Refs.~\cite{for18,pod06}, respectively.}
\end{figure}

\end{appendix}

\end{document}